\documentclass[fleqn,usenatbib]{mnras}

\usepackage[T1]{fontenc}
\usepackage{ae,aecompl}

\usepackage{graphicx}	
\usepackage{amsmath}	
\usepackage{amssymb}	

\usepackage{newtxtext,newtxmath}
\usepackage{mathptmx}
\usepackage{txfonts}

\usepackage{pdflscape}
\usepackage{longtable}
\usepackage{supertabular}

\usepackage{rotating}
\usepackage{footnote}
\usepackage{times}
\usepackage{chemmacros}

\usepackage{tikz,xcolor,hyperref}

\definecolor{lime}{HTML}{A6CE39}
\DeclareRobustCommand{\orcidicon}{
	\begin{tikzpicture}
	\draw[lime, fill=lime] (0,0) 
	circle [radius=0.16] 
	node[white] {{\fontfamily{qag}\selectfont \tiny ID}};
	\draw[white, fill=white] (-0.0625,0.095) 
	circle [radius=0.007];
	\end{tikzpicture}
	\hspace{-2mm}
}

\foreach \x in {A, ..., Z}{\expandafter\xdef\csname orcid\x\endcsname{\noexpand\href{https://orcid.org/\csname orcidauthor\x\endcsname}
			{\noexpand\orcidicon}}
}

\usepackage[normalem]{ulem}  
\newcommand\redout{\bgroup\markoverwith
{\textcolor{red}{\rule[0.5ex]{2pt}{0.8pt}}}\ULon}

\newcommand{\ha}{H\,$\alpha$} 
\newcommand{\hb}{H\,$\beta$}

\newcommand{\argoniii}{[Ar\,{\sc iii}]}

\newcommand{\heliumb}{He\,{\sc ii}}
\newcommand{\ariii}{[Ar\,{\sc iii}]~7136~\AA}
\newcommand{\nia}{[N\,{\sc i}]~5199~\AA}
\newcommand{\oi}{[O\,{\sc i}]~6300~\AA}
\newcommand{\oiia}{[O\,{\sc ii}]~7320~\AA}
\newcommand{\oiib}{[O\,{\sc ii}]~7330~\AA}
\newcommand{\niia}{[N\,{\sc ii}]~5755~\AA}
\newcommand{\niib}{[N\,{\sc ii}]~6584~\AA}
\newcommand{\niic}{[N\,{\sc ii}]~6548~\AA} 
\newcommand{\siia}{[S\,{\sc ii}]~6717~\AA}
\newcommand{\siib}{[S\,{\sc ii}]~6731~\AA}
\newcommand{\cliiia}{[Cl\,{\sc iii}]~5517~\AA}
\newcommand{\cliiib}{[Cl\,{\sc iii}]~5538~\AA} 

\newcommand{\oiiia}{[O\,{\sc iii}]~4959~\AA} 
\newcommand{\oiiib}{[O\,{\sc iii}]~5007~\AA}

\newcommand{\siiia}{[S\,{\sc iii}]~6312~\AA} 
\newcommand{\siiib}{[S\,{\sc iii}]~9069~\AA} 
\newcommand{\nitrogen}{[N\,{\sc ii}]}

\newcommand{\nitrogena}{[N\,{\sc i}]}
\newcommand{\oxygeniii}{[O\,{\sc iii}]}

\newcommand{\oxygeni}{[O\,{\sc i}]}
\newcommand{\oxygenii}{[O\,{\sc ii}]}
\newcommand{\oxygeniirec}{O\,{\sc ii}}
\newcommand{\carboniirec}{C\,{\sc ii}}
\newcommand{\nitrogeniirec}{N\,{\sc ii}}

\newcommand{\sulfur}{[S\,{\sc iii}]}
\newcommand{\sulfurt}{[S\,{\sc ii}]}
\newcommand{\chloro}{[Cl\,{\sc iii}]}

\newcommand{\carboni}{[C\,{\sc i}]}

\def\vhel{\ifmmode{V_{{\rm HEL}}}\else{$V_{{\rm HEL}}$}\fi}
\def\vsys{\ifmmode{V_{\rm sys}}\else{$V_{\rm sys}$}\fi}
\def\kms{\ifmmode{~{\rm km\,s}^{-1}}\else{~km~s$^{-1}$}\fi}
\def\vlsr{\ifmmode{v_{\rm lsr}}\else{$v_{\rm lsr}$}\fi}

\title[{\sc satellite}: NGC~7009 and NGC~6778 case studies]{Spectroscopic Analysis Tool for intEgraL fieLd unIt daTacubEs ({\sc satellite}): Case studies of NGC~7009 and NGC~6778 with MUSE}

\author[Stavros Akras]{S. Akras$^{1\orcidA{}}$~\thanks{E-mail: stavrosakras@gmail.com}, H. Monteiro$^{2\orcidB{}}$, J. R. Walsh$^{3\orcidC{}}$, J. Garc\'{i}a-Rojas$^{4,5\orcidD{}}$, I. Aleman$^{2\orcidE{}}$, 
\newauthor{H. Boffin$^{3\orcidF{}}$, P. Boumis$^{1\orcidG{}}$, A. Chiotellis$^{1\orcidH{}}$, R. M. L. Corradi$^{5,6\orcidI{}}$, D. R. Gon\c{c}alves$^{7}$,}
\newauthor{L. A. Guti\'{e}rrez-Soto$^{8\orcidK{}}$, D. Jones$^{4,5\orcidL{}}$, C. Morisset$^{9\orcidM{}}$, X. Papanikolaou$^{10}$}
\\
$^{1}$Institute for Astronomy, Astrophysics, Space Application \& Remote Sensing, National Observatory Athens, GR-15236, Athens, Greece\\
$^{2}$Instituto de F\'isica e Qu\'imica, Universidade Federal de Itajub\'a, Av. BPS 1303, Pinheirinho, 37500-903, Itajub\'a, MG, Brazil\\
$^{3}$European Southern Observatory. Karl-Schwarzschild Strasse 2, D-85748 Garching, Germany\\
$^{4}$Instituto de Astrof\'{i}sica de Canarias (IAC), E-38205 La Laguna, Tenerife, Spain\\
$^{5}$Departamento de Astrof\'{i}sica, Universidad de La Laguna, E38206, La Laguna, Tenerife, Spain\\
$^{6}$GRANTECAN, Cuesta de San Jos\'{e} s/n, E-38712 Bre\~{n}a Baja, La Palma, Spain\\
$^{7}$Observat\'{o}rio do Valongo, Universidade Federal do Rio de Janeiro, Ladeira Pedro Antonio 43, Rio de Janeiro 20080-090, Brazil\\
$^{8}$Departamento de Astronomia, IAG, Universidade de S\~{a}o Paulo, Rua do Mat\~{a}o, 1226, 05509-900, S\~{a}o Paulo, Brazil\\
$^{9}$Universidad Nacional Aut\'{o}noma de M\'{e}xico, Instituto de Astronom\'{i}a, AP 106, Ensenada, 22800, BC, Mexico\\
$^{10}$Dionysos Satellite Observatory (DSO), School of Rural, Surveying and Geoinformatics Engineering, \\
National Technical University of Athens, Iroon Polytechneiou Str. 9, 15780 Zografou, Greece
}

\date{Accepted XXX. Received YYY; in original form ZZZ}

\pubyear{2020}

\begin{document}
\label{firstpage}
\pagerange{\pageref{firstpage}--\pageref{lastpage}}
\maketitle

\begin{abstract}
Integral field spectroscopy (IFS) provides a unique capability to spectroscopically study extended sources over a 2D field of view, but it also requires new techniques and tools. In this paper, we present an automatic code, Spectroscopic Analysis Tool for intEgraL fieLd unIt daTacubEs, {\sc satellite}, designed to fully explore such capability in the characterization of extended objects, such as planetary nebulae, H~{\sc ii} regions, galaxies, etc. {\sc satellite} carries out 1D and 2D spectroscopic analysis through a number of pseudo-slits that simulate slit spectrometry, as well as emission line imaging. The 1D analysis permits direct comparison of the integral field unit (IFU) data with previous studies based on long-slit spectroscopy, while the 2D analysis allows the exploration of physical properties in both spatial directions. Interstellar extinction, electron temperatures and densities, ionic abundances from collisionally excited lines, total elemental abundances and ionization correction factors are computed employing the {\sc pyneb} package. A Monte Carlo approach is implemented in the code to compute the uncertainties for all the physical parameters. {\sc satellite} provides a powerful tool to extract physical information from IFS observations in an automatic and user configurable way. The capabilities and performance of {\sc satellite} are demonstrated by means of a comparison between the results obtained from the Multi Unit Spectroscopic Explorer (MUSE) data of the planetary nebula NGC~7009 with the results obtained from long-slit and IFU data available in the literature. The {\sc satellite} characterization of NGC~6778 based on MUSE data is also presented.

\end{abstract}

\begin{keywords}
planetary nebulae: general -- planetary nebulae: individual: NGC~7009, NGC 6778 -- H II regions -- ISM: abundances -- methods: numerical; Astronomical instrumentation, methods, and techniques
\end{keywords}



\section{Introduction}

Recent advances in integral field spectroscopy (IFS) have promoted a huge growth in imaging spectroscopy, demanding new approaches to the study of extended and resolved objects. Over the last two decades several integral field units (IFU) have been built covering mainly optical and infrared wavelengths, with diverse characteristics, such as field of view, wavelength coverage, spatial resolution, resolving power~\citep[see conference review on IFS, ][]{Allington2006, Mediavilla2011}.

IFS provides data in 3 dimensions (2 spatial and 1 spectral) or, equivalently, a cube of data where each wavelength element corresponds to a spectral image. These data cubes make possible the simultaneous spectroscopic analysis of extended objects (e.g. galaxies, H~{\sc ii} regions, planetary nebulae) in both spatial directions providing the spatial distribution of emission lines fluxes, line ratios and the nebular physical parameters (e.g. extinction, electron temperature and density ($T_{e}$, $N_{e}$), abundances), a task that is very time-consuming with traditional long-slit spectroscopy techniques \citep[e.g.][]{Monteiro2004,Monteiro2005,Phillips2010,Ferreira2011} or CCD imaging spectroscopy \citep[e.g.][]{Jacoby1987,Lame1994,Lame1996}.

These advances have boosted the number of works devoted to spatially resolved studies of planetary nebulae (PNe) using IFS over the last decade; some representative examples of these studies are, e.~g. \citet{Tsamis2008}, \citet{Monteiro2013}, \citet{Ali2016}, \citet{Walsh2016}, \citet{Walsh2018}, \citet{Ali2019}, \citet{Monreal2020}, \citet{Akras2020a}, and \citet{Garciarojas2022}.

The vast majority of spectroscopic studies of extended objects have been performed employing the traditional long-slit spectroscopy. As a consequence, spectroscopic data are available only for specific regions where the slits or apertures were placed. Moreover, diagnostic diagrams such as BPT \citep{Baldwin1981}, VO \citep{Veilleux1987} and STB \citep{Sabbadin1977}, built to distinguish AGN, LINERs, and Seyfert galaxies or PNe, SNRs and H~{\sc ii} regions are also based on 1D long-slit data as well as simulations from 1D models such as {\sc cloudy} \citep{Ferland2013,Ferland2017} and {\sc mappings} \citep{Sutherland2017,Sutherland2018}.

From IFU data, besides the unique 2D spectroscopic analysis, it is also possible to perform 1D analysis by simulating slit apertures (hereafter \lq\lq pseudo-slits\rq\rq). This is essential to acquire results that can be properly compared with those from long-slit studies present in the literature. For instance, it has already been pointed out that a spaxel-by-spaxel analysis of emission line ratios of extended sources from IFUs is not recommended for a straightforward comparison with integrated long-slit spectra and the resultant diagnostic diagrams \citep{Ercolano2012,Morisset2018,Akras2020a}.

In this paper, we present a newly developed automatic code that performs a full spectroscopic analysis of extended sources using IFS data, namely \lq\lq Spectroscopic Analysis Tool for intEgraL fieLd unIt daTacubEs ({\sc satellite})\rq\rq. The novelty of this tool is that it provides an 1D spectroscopic analysis through pseudo-slits and 2D analysis through maps. The former will allow to properly compare the results obtained from IFUs with those from previous 1D long-slit spectroscopy while the latter will allow to explore the extended sources in both spatial directions. A brief presentation of the capabilities of {\sc satellite} is presented in \cite{Akras2020a} using VIMOS IFS data of the PN Abell~14.

The paper is organized as follows. In Section 2, we present the four modules of the {\sc satellite} code available in the current version namely: {\it (I) rotation analysis, (II) radial analysis, (III) specific slits analysis and (IV) 2D analysis}. In Sections~3 and 4, we apply the {\sc satellite} code to the Multi Unit Spectroscopic Explorer (MUSE) Science Verification data of the PN NGC~7009, as well as the data from program 097.D-0241(A) (PI: Corradi). A comparison between {\sc satellite's} results and those from \cite{Walsh2018} is presented as well as the outcomes from the two MUSE datacubes. In Section~5, we apply the {\sc satellite} code to the MUSE data of NGC~6778 and perform the 1D and 2D spectroscopic characterization of the nebula. The results of this work as well as the potential upgrades to future versions of the {\sc satellite} code are discussed in Section~6.

\section{The {\sc satellite} code}

The Spectroscopic Analysis Tool for intEgraL fieLd unIt daTacubEs ({\sc satellite}) is a newly developed {\sc python 3} code that performs a number of spectroscopic analyses on IFU data. As input, {\sc satellite} uses the flux maps of several emission lines usually detected in galaxies, H~{\sc ii} regions and PNe\footnote{The analysis of weak collisionally excited lines and recombination lines from O or N is not implemented yet.}, extracted from IFU datacubes. 

{\sc satellite} simulates pseudo-slit spectra by summing up the values of each individual spaxel (the unit of the IFU) within the specific region defined by the width and length of the pseudo-slits provided by the user. Note that the pseudo-slits are built considering full-size spaxels and no partially covered ones. 

To exclude values from problematic spaxels or with low signal-to-noise (S/N) ratio, two criteria must be satisfied by each spaxel: (i) F(\ha)$>$0, F(\hb)$>$0 and (ii) F(\ha)$>$2.85*F(\hb), to avoid unrealistic values for the interstellar extinction coefficient (c(\hb))\footnote{Negative c(\hb) values are still possible especially in the halo of PNe due to scattering process. \cite{Walsh2018} reported negative c(\hb) values at the halo of NGC~7009 after applying the Voronoi tesselation method. This technique is not yet implemented in the {\sc satellite} code.}.

\begin{figure}
\centering
\includegraphics[width=7.25cm]{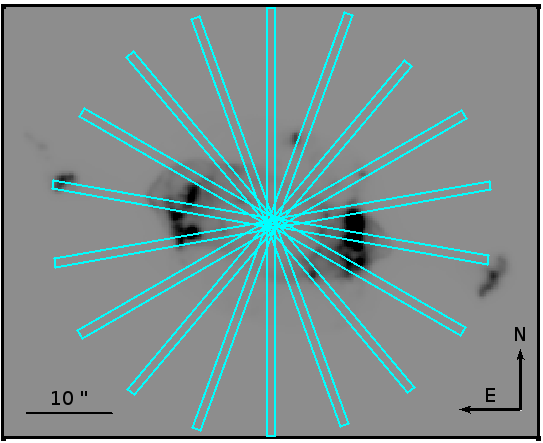}
\caption{Illustrative example of pseudo-slits positions in NGC~7009 overlaid on the \nitrogen~6584\AA\ image for the rotation analysis module. The width, length, initial, final and step angle are provided by the user. The size of the image is 62$\times$50 arcsec.}
\label{figrotate}
\end{figure}

\begin{figure}
\centering
\includegraphics[width=8cm]{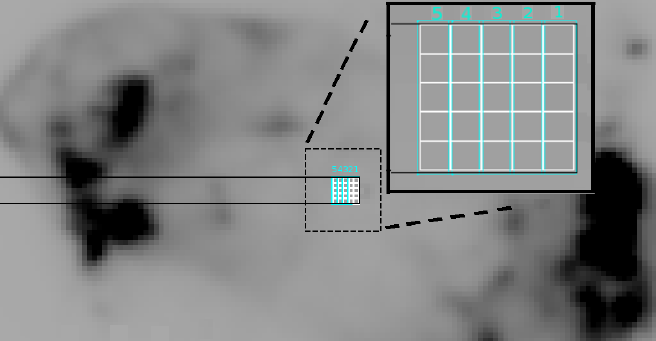}
\caption{Illustrative example of the {\it radial analysis} module for the PA=90 degrees pseudo-slit in NGC~7009 overlaid on the \nitrogen~6584\AA\ image. The inset image shows a zoomed-in of the pseudo-slit. The boxes correspond to the spaxels of the IFU. The fluxes of the spaxels in each numbered column/row are summed up to get the total flux for each column/row within the width of the pseudo-slit or equivalently to the distance from the central star taking into account the pixel scale of the IFU.}
\label{figradialslit}
\end{figure}

For all the modules, {\sc satellite}\footnote{{\sc satellite} is available to the GitHub website with its proper documentation and example {https://github.com/StavrosAkras/SATELLITE.git}.} first calculates c(\hb) considering an extinction curve law selected by the user among the options in the {\sc pyneb} package version 1.1.15 \citep{Luridiana2015}. Then it computes the physical parameters of the ionized nebula such as $T_{e}$ and $N_{e}$, ionic abundances, ionization correction factors (ICFs), total elemental abundances and abundance ratios,  employing the {\sc pyneb} package. In this work, we used the default atomic databases in the {\sc pyneb} package \citep[see][]{Morisset2020}. The diagnostic lines for the determination of $T_{e}$ and $N_{e}$ as well as the values applied for the ionic abundances determinations are set by the user from a list of diagnostics. The uncertainties of the physical parameters are computed following Monte Carlo simulations. The number of random spectra (hereafter replicate spectra) is chosen by the user and they are generated considering a Gaussian distribution centred at a reference spectrum (i.e. line intensities) with a sigma equal to the uncertainty of each line intensity computed from the provided error maps\footnote{An additional error can also be considered for each emission line as a percentage of the line flux for all the systematic uncertainties. This error is a free parameter in the code and it is given by the user.}. Then, all the physical parameters are computed for each replicate spectrum and the resultant standard deviation of the distribution of each parameter is considered to be the final uncertainty provided by the {\sc satellite} code. Besides the calculations of nebular physical parameters for the pseudo-slits, {\sc satellite} also provides scatter plots, emission line diagnostic diagrams, histograms and 2D maps for all nebular parameter. The modules of {\sc satellite} are described below.

\subsection{Rotation analysis module}
This module deals with the spectroscopic analysis of an extended source from a number of pseudo-slits placed radially across half of the nebula with position angles (PAs) from 0 to 360 degrees (Fig.~\ref{figrotate}). {\sc satellite} first rotates all raw images (line flux maps) and then generates new images from which it computes the fluxes summing up the values of each individual pixel with the pseudo-slit. The pseudo-slits are always in vertical direction (up-and-down orientation).

PA=0~degrees corresponds to a pseudo-slit in vertical orientation on the map and increases as the pseudo-slit rotates in the anti-clockwise direction. If the IFU data and the line flux maps are oriented such that North is up and East is left, the PA of the pseudo-slits in the {\sc satellite's} frame coincides with the slits's PA on-sky values (PA=0~degrees in the North-South direction and increases rotating East from North). The initial, final and step angles of the position angle as well as the width and the length of the pseudo-slits are given by the user. 

All the line intensities and physical parameters of the nebula are derived for each pseudo-slit. Therefore, this module provides an analysis of nebular gas (c(\hb), $T_{e}$ and $N_{e}$, ionic and elemental abundances, ICFs) as functions of slit PA as well as an ASCII file (a multi-column list of data used for the plots) so that users can construct their own plots.

\subsection{Radial analysis module}
The {\it radial analysis} module computes the emission line fluxes along a specific pseudo-slit as a function of the distance from the geometric center and/or the ionizing source (e.g. central star). The characteristics of this pseudo-slit (PA, width and length) are free parameters provided by the user. {\sc satellite} calculates the line fluxes summing up the values of the individual spaxels in one column/row enclosed by the pseudo-slit (Fig.~\ref{figradialslit}) and then all the physical parameters of the nebula at each column/row). Scatter plots for each parameter as a function of the distance taking into account the pixel scale of the IFU are provided as well as an ascii file with the data. The distance from the central star at which each emission line shows a peak is also determined. The emission lines for this task are also chosen by the user. Overall, the {\it radial analysis} module makes possible the study of nebular parameters and line ratios as a function of the distance from the central ionising source of the nebula.

\subsection{Specific slits analysis module}
The {\it specific slits analysis} module performs a full spectroscopic analysis for 10 (default number in {\sc satellite}~v1.3) distinct regions that can be configured so as to study specific features/regions that possess distinctive morphological or physical properties, such as lobes, ansae, low-ionization structures (LISs), rim, etc. in PNe for a direct comparison with the results from long-slit studies.

\begin{figure}
\centering
\includegraphics[width=7.5cm]{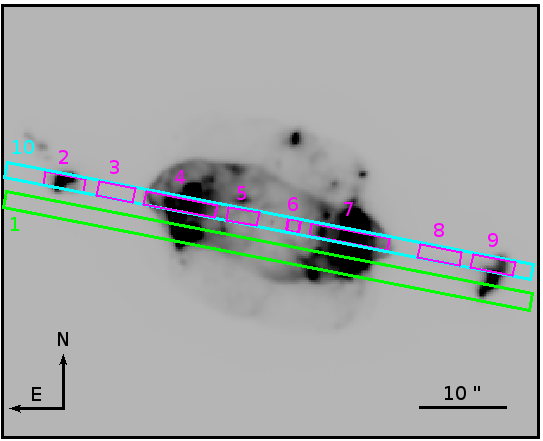}
\caption{Ten selected regions in NGC~7009 overlaid on the \nitrogen~6584\AA\ image. The position of the centre (x, y coordinates), position angle, width and length of the slits are free parameters provided by the user. Slits 1 and 10 represent the slits position from \protect\cite{Fang2011} and \protect\cite{Goncalves2003}, respectively. Numbered regions from 2 to 9 correspond to the sub-structures of knots and jet-like, or sub-regions of rims defined in \protect\cite{Goncalves2003}. The size of the image is 62$\times$50 arcsec.}
\label{figregions}
\end{figure}

Line intensities and physical parameters are calculated for each specific region defined by the pseudo-slits summing up all the spaxels values. The exact position/centre, PA, width and length of the pseudo-slits are not necessarily the same and are provided by the user. Fig.~\ref{figregions}~displays the 10 selected regions for the analysis of NGC~7009 overlaid on its \nitrogen~6584\AA\ image.

\begin{table*}
\caption{Comparison between {\sc satellite} and \protect\cite{Walsh2018} results obtained from the NGC~7009 Science Verification MUSE data. Flux maps used for this comparison were those extracted by \protect\citet{Walsh2018}.} \label{tab:comparison}

\begin{tabular}{lcccccccc}

\hline
Parameter &	{\sc satellite} & SD$^\dag$   & \cite{Walsh2018} & SD	& Log(ratio) & SD  & \# of spaxels\\
\hline
c(\hb)	  & 0.105	        & 0.065	&	0.122	          &	0.041	&	-0.022 & 0.165	& 28070\\
Te(SIII6312\_9069)\_Ne(ClIII5517\_5538)	  & 9166	        & 301 	& 9159	          &	326	&	0.0005 & 0.0024 & 17305\\
Ne(ClIII5517\_5538)\_Te(SIII6312\_9069)	  & 3609	        & 2037	&	3547	          &	2476	&	0.0001 & 0.0015 & 17305\\
\hline	  
log(He~{\sc i}~ 5876/\ha) & -1.264 & 0.035 &	-1.270	& 0.031 & -0.004 & 0.002 & 35876	\\
log(He~{\sc ii}~ 5412/\hb) & -2.443 &	0.484 &	-2.425	& 0.484 & 0.004  & 0.001 & 14427\\
log([N~II] 6583/\ha)& -1.465 & 0.379 &  -1.447	& 0.336 & 0.0001 & 0.0014 & 39302\\
log([O~I] 6300/\ha) & -2.953 & 0.833 &	-2.880	& 0.863 & -0.0009 & 0.0007 & 11720\\	
log([O~{\sc iii}] 4959/\hb)& 0.666 & 0.080 & 0.699   & 0.078 & -0.0025 & 0.0016 & 40724\\	  
\hline	  
\ch{He+} (5876)/\ch{H+}  & 0.105 & 0.008 & 0.102$^{\dag\dag}$ & 0.009 & 0.013 & 0.012 & 16516\\
\ch{He+} (6678)/\ch{H+}  & 0.100 & 0.008 & 0.102$^{\dag\dag}$ & 0.009 & -0.010 & 0.010 & 16516\\
\ch{He^{++}} (5412)/\ch{H+} & 0.007 & 0.008 & 0.007 & 0.008 & -0.001 & 0.003 & 14042\\	  
N+ (6548)/\ch{H+} (10$^{-6}$) & 5.688 & 14.499 & 5.875$^{\dag\dag}$ & 14.66 & -0.017 & 0.014 & 17304\\
N+ (6584)/\ch{H+} (10$^{-6}$) & 5.789 & 14.752 & 5.875$^{\dag\dag}$ & 14.66 & -0.009 & 0.014 & 17304\\	 
\ch{O+} (7320)/\ch{H+} (10$^{-5}$) & 3.491 & 3.384 & 3.068 & 2.703 & 0.060 & 0.031 & 17296\\
\ch{O+} (7330)/\ch{H+} (10$^{-5}$) & 3.728 & 3.351 & 3.068 & 2.703 & 0.090 & 0.028 & 17283 \\
\ch{O^{++}} (4959)/\ch{H+} (10$^{-4}$) & 5.659& 0.688 & 5.992 & 0.765 & -0.024 & 0.011  & 17253\\
\ch{S+} (6716)/\ch{H+} (10$^{-7}$) & 3.176 & 6.882 & 2.977$^{\dag\dag}$ & 6.101 & 0.014 & 0.055 & 17301\\
\ch{S+} (6731)/\ch{H+} (10$^{-7}$) & 2.927 & 6.102 & 2.977$^{\dag\dag}$ & 6.101 & -0.010 & 0.014 & 17305\\
\ch{S^{++}} (6312)/\ch{H+} (10$^{-6}$) & 4.490 & 1.548 & 4.616 & 1.618 & -0.012 & 0.017 & 17305\\	 
\ch{Cl^{++}} (5517)/\ch{H+} (10$^{-7}$) & 0.998 & 0.240 & 1.019 & 0.248 & -0.009 & 0.012 & 17305\\
\ch{Ar^{++}} (7136)/H+ (10$^{-6}$) & 1.645 & 0.337 & 1.837 & 0.384 & -0.048 & 0.015 & 17305\\	 
\hline
\end{tabular}
\begin{flushleft}
$^{\dag}$ Standard Deviation from the whole map. $^{\dag\dag}$ These are average ionic abundances from
both ions.
\citep{Walsh2018} data are available in the 
The Strasbourg astronomical Data Center (http://cdsarc.u-strasbg.fr/viz-bin/qcat?J/A+A/620/A169)
\end{flushleft}
\end{table*}

\begin{figure*}
\centering
\includegraphics[width=7.5cm]{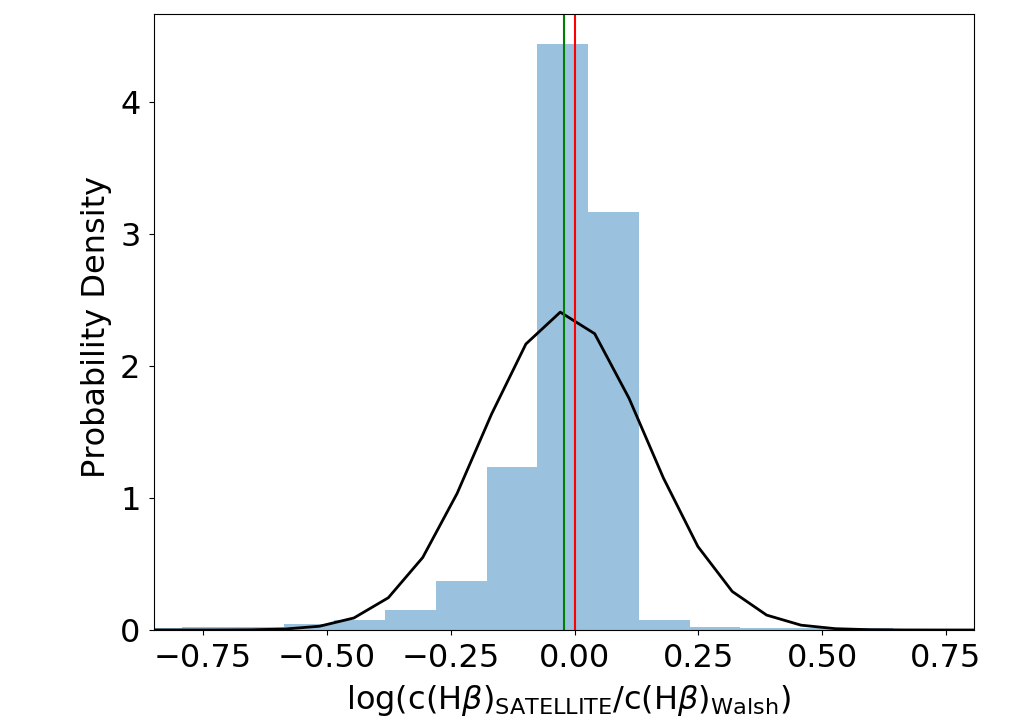}
\includegraphics[width=7.5cm]{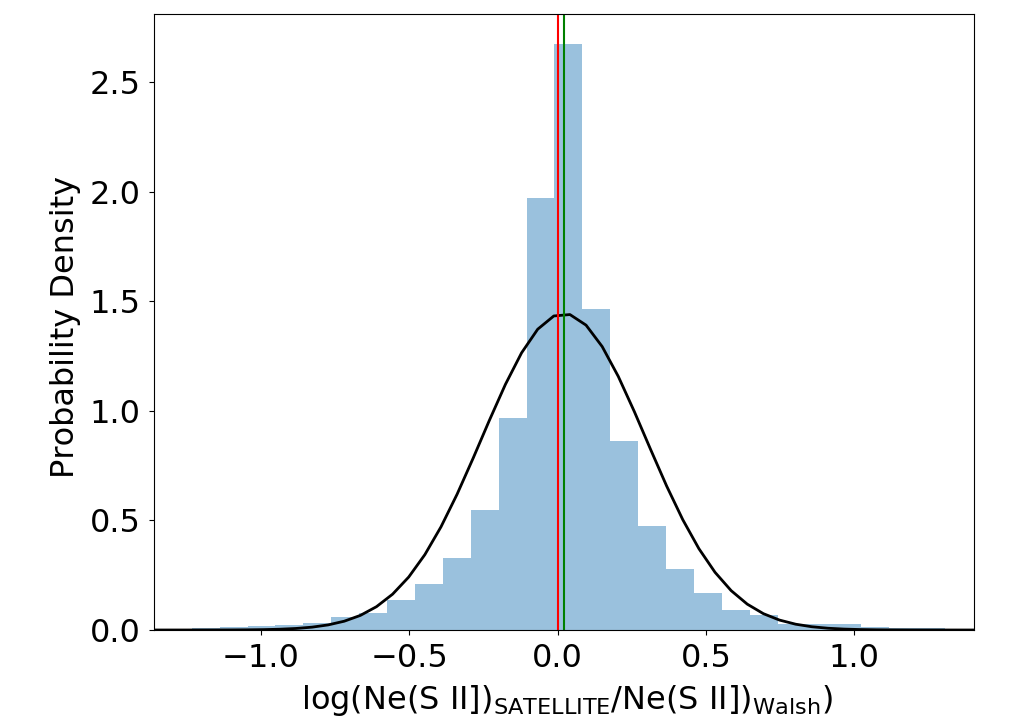}
\includegraphics[width=7.5cm]{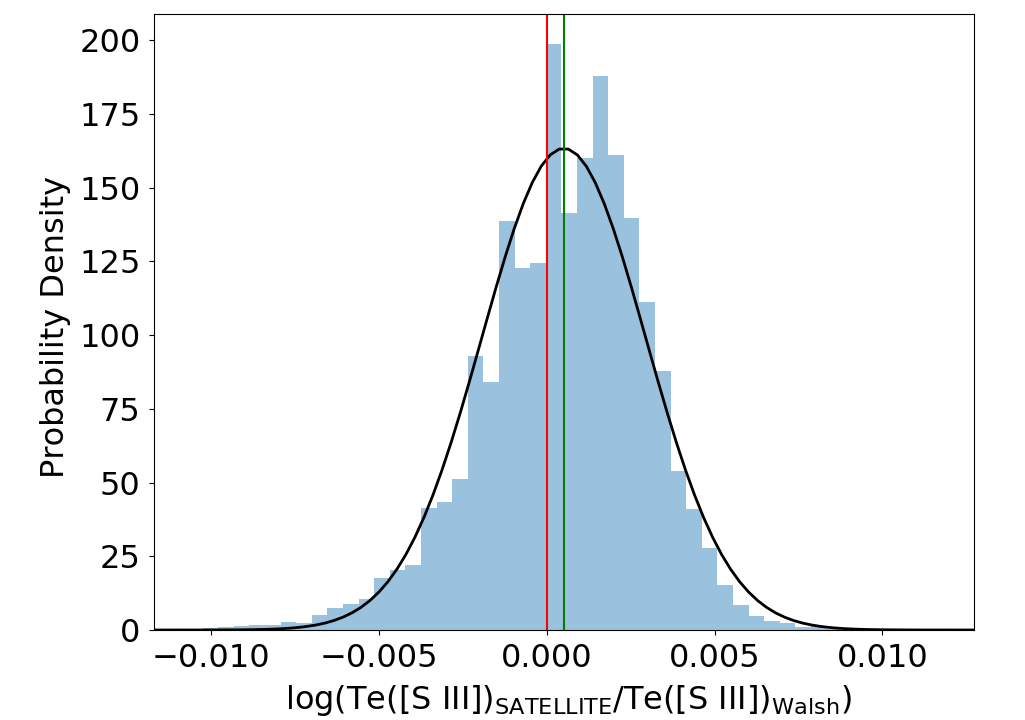}
\includegraphics[width=7.5cm]{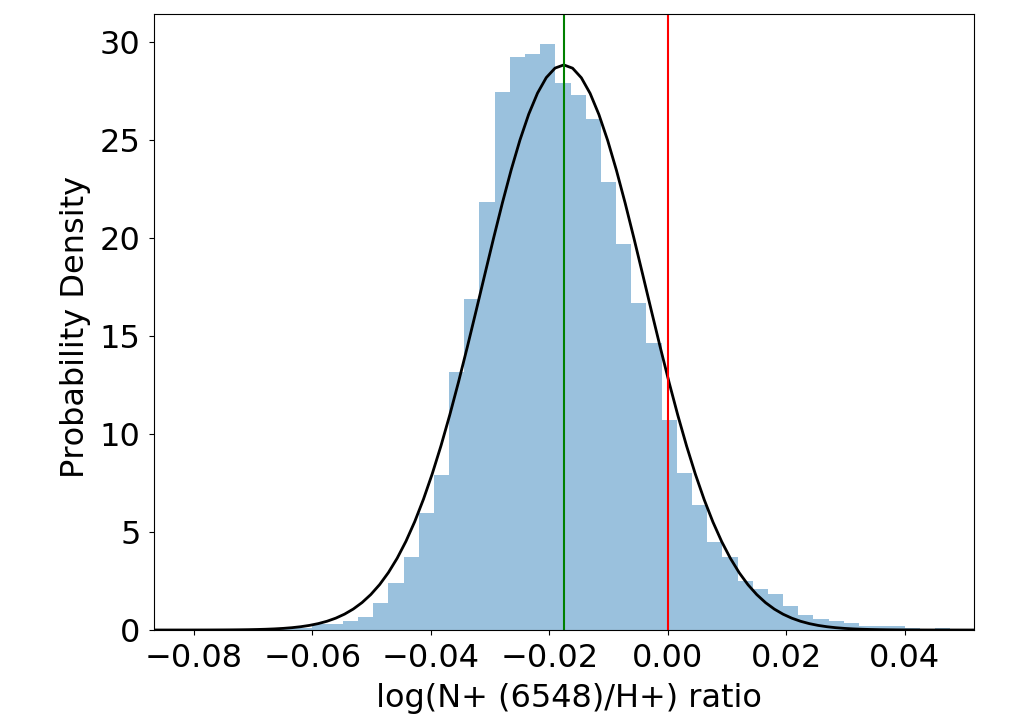}
\caption{Comparison distributions of the log(c(\hb)$_{\rm{\sc satellite}}$ / c(\hb)$_{\rm{Walsh}}$) (upper, left panel), log($N_{e}$\sulfurt$_{\rm{\sc satellite}}$ / $N_{e}$\sulfurt$_{\rm{Walsh}}$) (upper right panel), log($T_{e}$\sulfur$_{\rm{\sc satellite}}$ / $T_{e}$\sulfur$_{\rm{Walsh}}$) (lower left panel) and log(\ch{N^+}(6584)/\ch{H^+}$_{\rm{\sc satellite}}$ / \ch{N^+}(6584)/\ch{H^+}$_{\rm{Walsh}}$) ratios (lower right panel) between the results obtained with {\sc satellite} and computed by \protect\cite{Walsh2018}. Red and green vertical lines indicate the zero value and the peak of each distribution. The black lines represent the probability density functions of the data.}
\label{figchb_TeNe_hist}
\end{figure*}

\subsection{2D analysis module}
The fourth {\sc satellite} module performs the full 2D spectroscopic analysis. This module calculates the line intensities and physical parameters in each spaxel and only for those that satisfy the aforementioned criteria. The principal outputs of this module are 2D  maps of line ratios, nebular parameters, ionic and total elemental abundances as well as the corresponding histograms of their distributions. Moreover, the mean values, standard deviations and percentiles of 5\%, 25\% (Q1), 50\% (median), 75\% (Q3), 95\% are also calculated for all the line ratios and nebular parameters.

A number of emission line diagnostic diagrams (including the classical BPT/VO, STB) selected by the user among a pre-defined list in {\sc satellite} are also generated and provided. There is an option to plot the values obtained from the {\it rotation analysis} and {\it specific slits analysis} modules and the {\it 2D analysis} from all spaxels on the same diagnostic diagrams.

The comparison of the emission line ratios from individual spaxel line ratios with the ratios obtained from integrated slits is crucial for the study of extended sources such as galaxies, H~{\sc ii} regions and PNe and the interpretation of the excitation mechanisms \citep{Ercolano2012,Morisset2018,Akras2020a}. Moreover, the available Ionization Correction Factors (ICFs) formulae for the computation of the total elemental abundances can be made only for 1D spectra rather than for each individual spaxel. The ionization structure of PNe is strongly dependent on the distance from the UV-source (e.g. PN central star) which can lead to misleading 2D abundances estimations. 

\begin{figure}
\centering
\includegraphics[width=7.8cm]{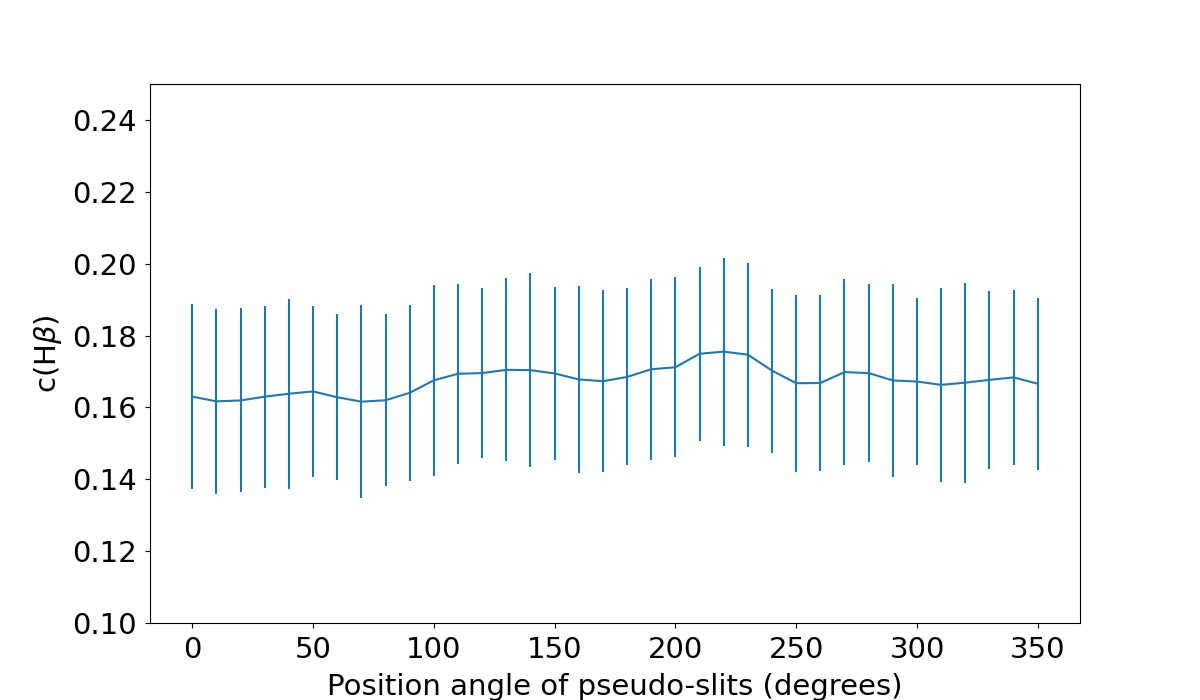}
\includegraphics[width=7.8cm]{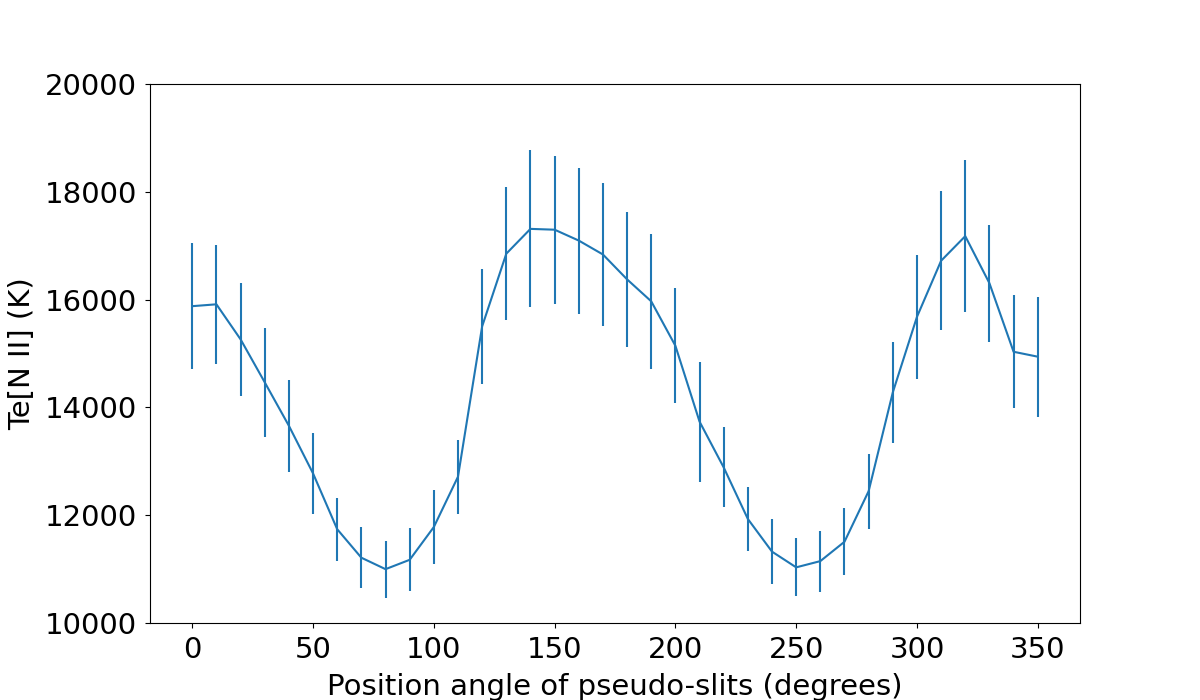}
\includegraphics[width=7.8cm]{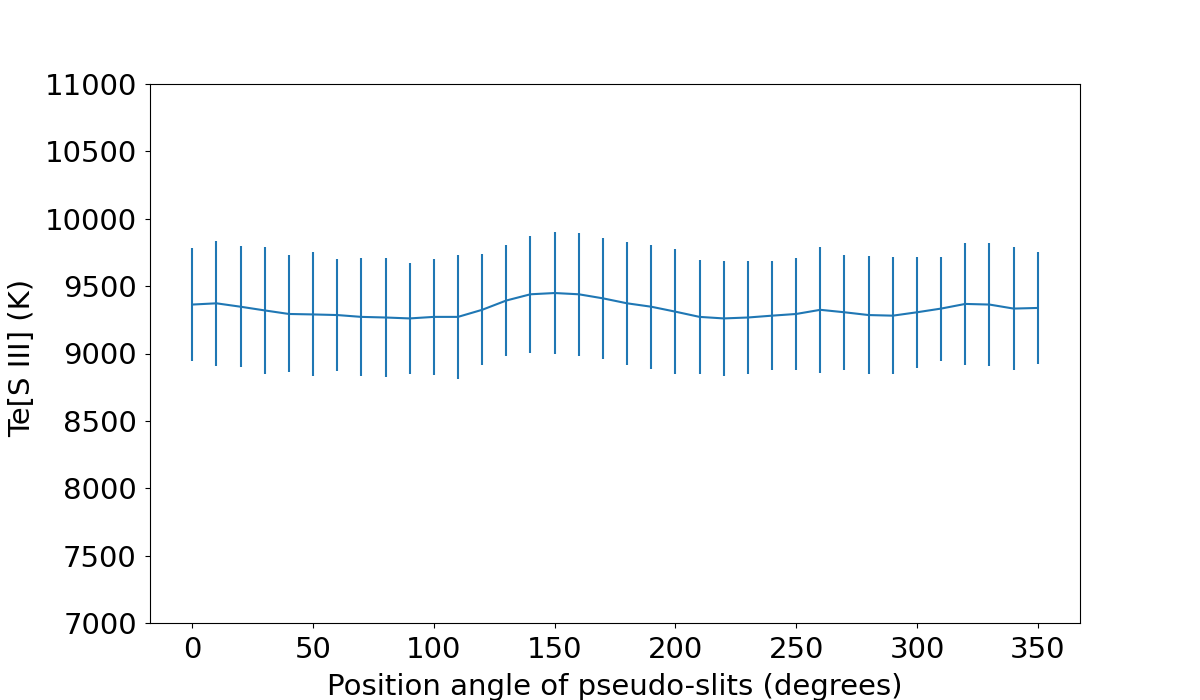}
\includegraphics[width=7.8cm]{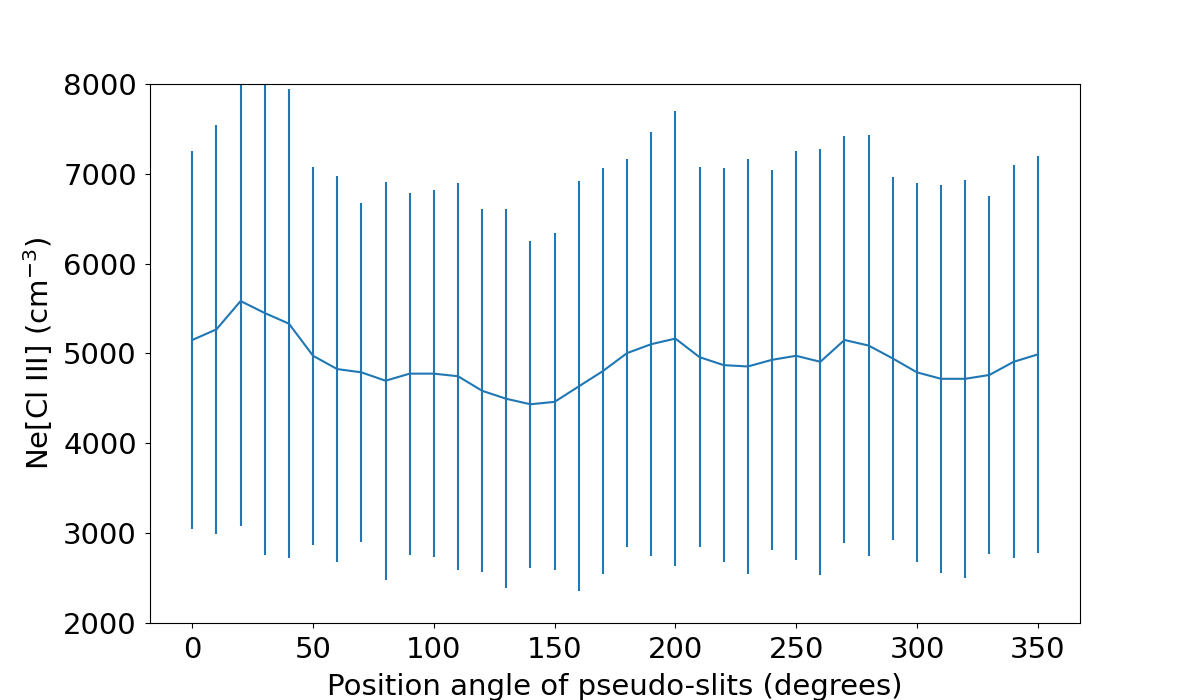}
\caption{c(\hb) (first panel), $T_{e}$\nitrogen (second panel), $T_{e}$\sulfur (third panel), and $N_{e}$\chloro (fourth panel)~versus the position angle of the pseudo-sits. PA=0 degrees corresponds to a pseudo-slit oriented north-south and increases rotating to the east.}
\label{figangles}
\end{figure}

\section{Case study of NGC~7009 with the Science Verification MUSE data}
In order to illustrate further and assess the numerical tools provided by {\sc satellite}, we use as a case study the planetary nebula NGC~7009, also known as the Saturn nebula, and it is among the most extensively studied PNe. Observations covering a large portion of the electromagnetic spectrum are available. Its high surface brightness has made possible various thorough studies using both imaging and spectroscopic data
\citep[e.g.][]{Lame1996,Guerrero2002,Rubin2002,Goncalves2003,Sabbadin2004,Goncalves2006,Rodriquez2007,Steffen2009,Phillips2010,Fang2011,Fang2013,Akras2020b}. As a consequence, NGC~7009 was also an ideal object for MUSE \citep{Bacon2010} Science Verification (SV) phase, and the first results from these data have been published by \cite{Walsh2016,Walsh2018}. 

The same emission line flux maps extracted from the SV MUSE datacube \citep{Walsh2018} available in VizieR On-line Data Catalog(J/ApJ/889/49) are used to test the performance of the {\sc satellite} code and compare its outputs with those from \cite{Walsh2018} as well as the long-slit spectra from \cite{Fang2011} and \cite{Goncalves2003}. The resulting maps from {\sc satellite} and \cite{Walsh2018} are first compared through a spaxel-by-spaxel approach. For this exercise, the same Galactic extinction law from \cite{Seaton1979} was considered for the spectroscopic analysis with {\sc satellite}. Table~1 lists the mean values and standard deviations for the maps of some physical parameters, logarithmic line ratios and ionic abundances. For the majority of the parameters the difference is found to be less than 5 percent with comparable dispersion (similar standard deviations). Note that the current version of the {\sc satellite} code does not correct for the contribution of recombination lines which may not be negligible, especially in PNe with relatively large abundance discrepancy factors between optical recombination lines and collisionaly excited lines, where this correction can be extremely problematic \citep[see ][]{GomezLlanos2020}. This explains the difference in the abundance of singly ionized Oxygen between {\sc satellite} and \citep{Walsh2018} as the \ch{O^{++}} recombination contribution results in 30~percent higher \ch{O^+}/\ch{H+}. The $T_{e}$\sulfur~and $N_{e}$\chloro~diagnostic line were considered for all the calculations in order to replicate the exact outputs from \cite{Walsh2018}. 

Fig.~\ref{figchb_TeNe_hist} displays some representative examples of comparison histograms of different parameters and abundance ratios obtained from the {\sc satellite} code and \cite{Walsh2018}. Probability density functions are also shown (black line) with peaks very close to zero, which verifies the consistency and robustness of the code.

\begin{figure}
\centering
\includegraphics[width=7.8cm]{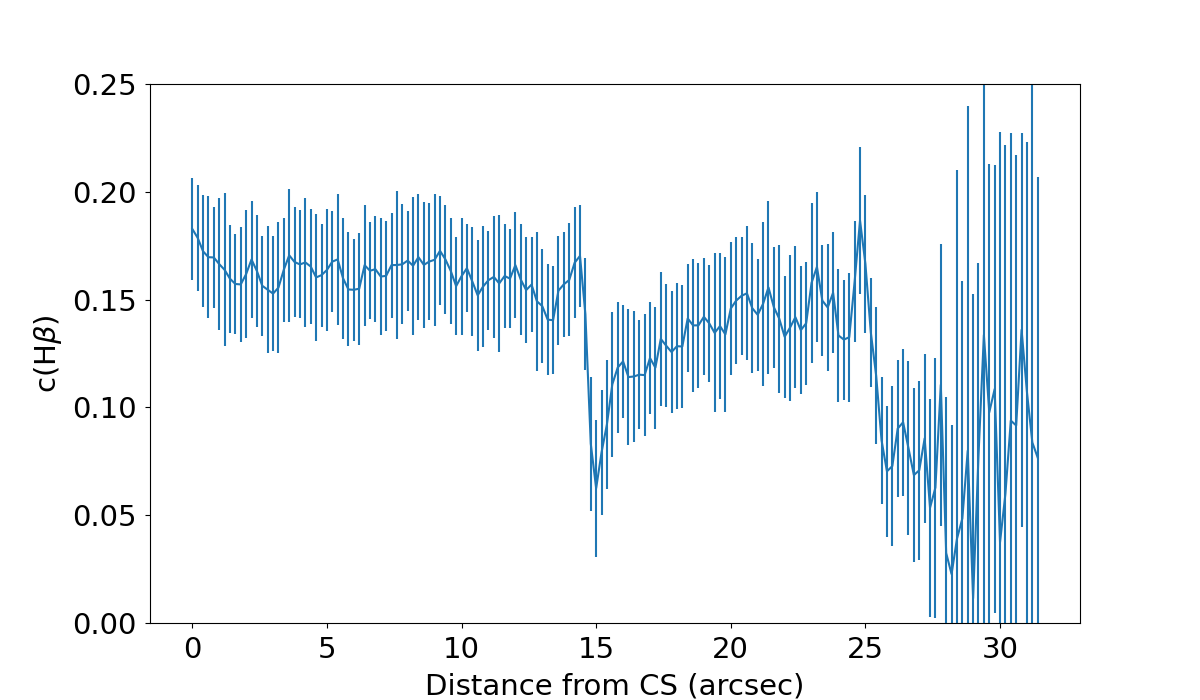}
\includegraphics[width=7.8cm]{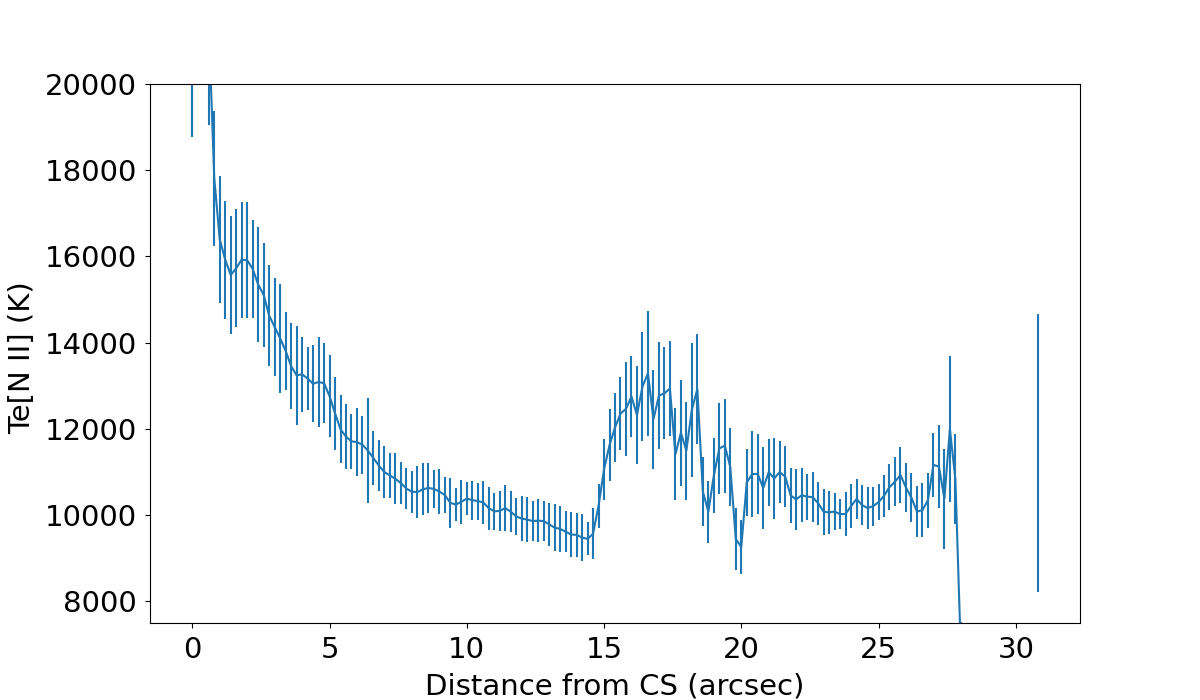}
\includegraphics[width=7.8cm]{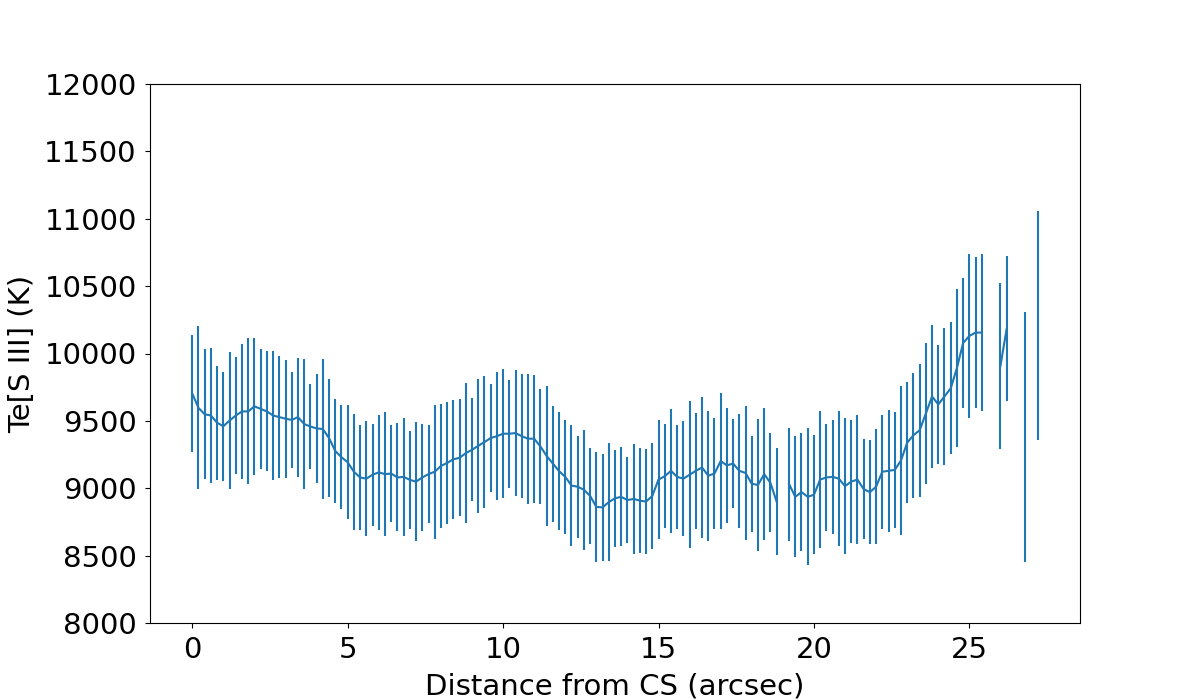}
\includegraphics[width=7.8cm]{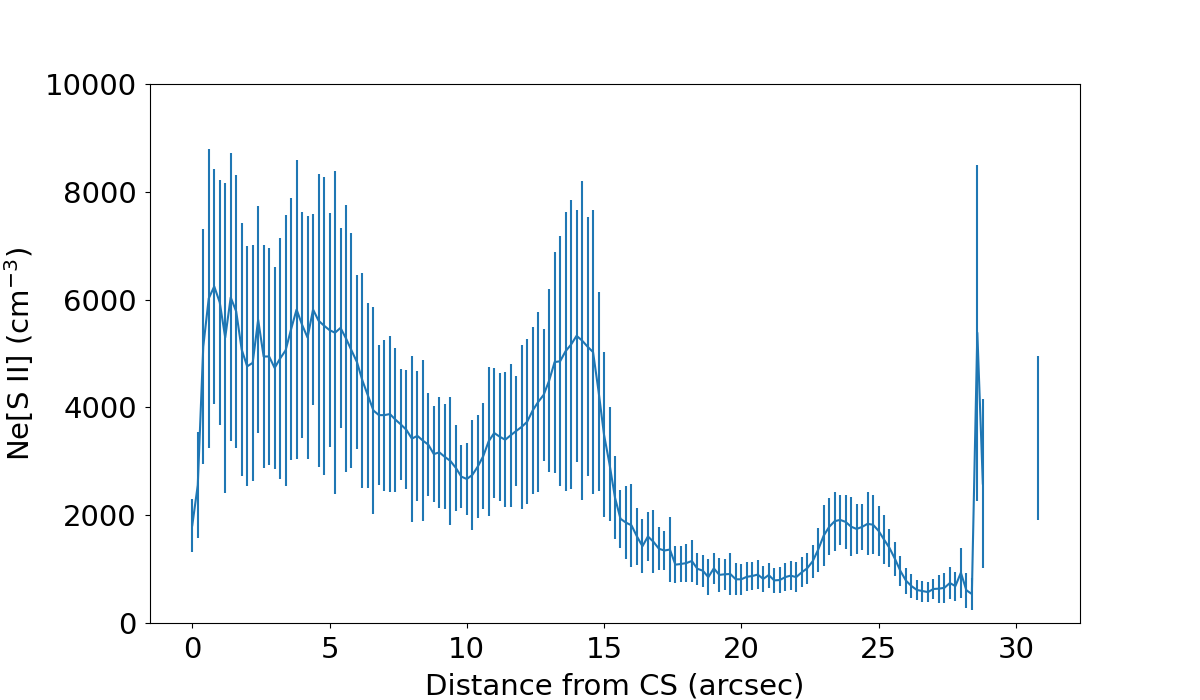}
\caption{c(\hb) (first panel), $T_{e}$\nitrogen~(second panel), $T_{e}$\sulfur~(third panel) and $N_{e}$\sulfurt~(fourth panel) as a function of the distance from the central star (in arcsec) in a pseudo-slit at PA=79 degrees.}
\label{figradial}
\end{figure}

\subsection{Rotation analysis module}

The complex morphology of NGC~7009 composed of the ellipsoidal main nebula, several sets of outer shells, loops, jet-like features and pairs of LISs, make it an ideal nebula for demonstrative purposes of {\sc satellite's} capabilities and the {\it rotation analysis} module. The position angle of the pseudo-slits varies from 0 to 360 with a step of 10 degrees. Fig.~\ref{figangles} illustrates the variation of c(\hb) (first panel), $T_{e}$\nitrogen~(second panel), $T_{e}$\sulfur~(third panel) and $N_{e}$\chloro~(forth panel) as functions of the PA.

\begin{figure}
\centering
\includegraphics[width=8.25cm]{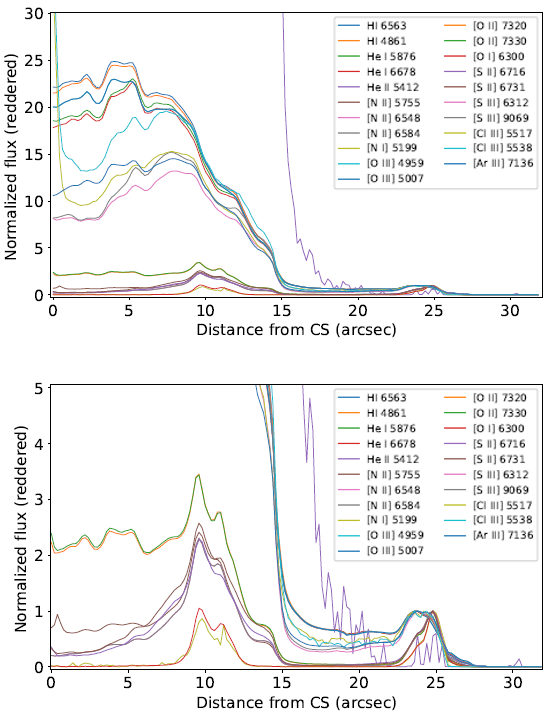}
\caption{Radial distribution of emission lines normalized to 1. The normalization is made at the peak values for distances $>$20 arcsec. Lower panel zooms in to better illustrate the distribution of weaker lines.}
\label{figradiallines}
\end{figure} 

c(\hb) slightly varies with the PA from 0.16 to 0.17  but it can be considered constant within the uncertainties. Thus, no divergence along the direction of the jet-like/LISs is found. $T_{e}$\nitrogen~shows strong variation with the PA between 10000~K up to 17000~K. However, this variation shall not be considered real. It is associated with the high values in the central region of the nebula reported by \cite{Walsh2018}. On the other hand, $T_{e}$\sulfur~displays a constant value of $\sim$9400~K, being 2000~K lower than the values estimated by \cite{Fang2011} despite the good agreement in line intensities. It should be noted that \citet{Fang2011} measured $T_{e}$ from the $T_{e}$\sulfur~6312/(9069+9530) line ratio while {\sc satellite} considers the $T_{e}$\sulfur~6312/9069 ratio\footnote{The location of the telluric features in the MUSE spectra of NGC 7009 (corrected for the motion of the Sun and Earth) was also verified without evidence for telluric features around 9069\AA~that could be responsible for the diminish of the line.}. From the reported intensities of the nebular \sulfur\ lines by \citet{Fang2011}, \sulfur~$\lambda$9530 line is likely dimmed by atmospheric absorption bands (the \sulfur~$\lambda$9531/$\lambda$9069 ratio is 1.88 in contrast with the expected theoretical value of 2.48 from the transition probabilities of \citealt{Mendoza1982}) resulting in higher temperature. This disagreement in $T_{e}$\sulfur~ is also responsible for the difference in the ionic abundances showed in Table~1. Nevertheless, the results from the Cor MUSE data ((097.D-0241(A), PI: R. L. M. Corradi; see Section~\ref{Cordata}) agree with the results from the earlier MUSE SV data. $N_{e}$\chloro~ can also be considered constant within the uncertainties. To avoid possible false variations in the chemical abundances of the nebula, $T_{e}$\sulfur~and $N_{e}$\chloro~were used to derive the chemical abundances and replicate the results from \cite{Walsh2018}.

\begin{table}
\caption{Distances of the outer emission line peak from the central star in a pseudo-slit at PA=79 degrees.} \label{distancepeak}
\begin{tabular}{cccc}
\hline
Line             & Distance$^{\dag}$ & Line & Distance\\
                 & (arcsec) &      & (arcsec)\\
\hline
H~{\sc i}~4861\AA\   & 23.6	& \niib                & 24.8 \\
\oiiia               & 23.8 & H~{\sc i}~6563\AA\   & 23.6 \\
\nia                 & 24.8 & \niic`               & 24.8 \\
He~{\sc ii}~5412\AA\ & 20.2 & He~{\sc i}~6678\AA\  & 23.8 \\
\cliiia              & 24.2 & \siia                & 24.8\\
\cliiib              & 24.4 & \siib                & 24.8\\
\niia                & 24.2 & \ariii               & 24.8\\
He~{\sc i}~5876\AA\  & 23.6 & \oiia                & 24.8\\
\oi                  & 24.8 & \oiib                & 24.8 \\
\siiia               & 23.8 & \siiib               & 23.8 \\
\hline
\end{tabular}
\begin{flushleft}
$^{\dag}$ The spacial resolution of MUSE maps is 0.2~arcsec.
\end{flushleft}
\end{table}

\subsection{Radial analysis module}
In Fig.~\ref{figradial}, we present the variation of c(\hb) (first panel), $T_{e}$\nitrogen~(second panel), $T_{e}$\sulfur~(third panel) and $N_{e}$\sulfurt~(fourth panel) as functions of the distance from the central star for a slit at PA=79 degrees (along the direction of the eastern jet/knot). c(\hb) is nearly flat ($\sim$0.16) for a distance up to 14-15 arcsec from the central star where it drops sharply to 0.06. This position corresponds exactly to the end of the ellipsoidal structure and the beginning of the jet-like structure. This drop in c(H$\beta$) is also highlighted in \cite{Walsh2016}. Then it gradually increases from 0.06 up to 0.16 at the distance of 25~arcsec. Above that distance, the diffuse and weaker \ha~and \hb~emission lines from the halo lead to large uncertainties and no robust results can be extracted.

\nitrogen~diagnostic lines yield a very high electron temperature in the inner nebula ($<$8~arcsec), which corresponds the the high value reported by \cite{Walsh2018}, and at a distance of 15-17~arcsec from the central star, where there is a peak of $\sim$14000~K. It is worth noting the radial distribution of $T_{e}$\sulfur. For distances up to $\sim$22 arcsec, $T_{e}$\sulfur~is nearly constant ($\sim$9200~K) and then it slightly increases to $\sim$10500K. This increase of $T_{e}$\sulfur~ may be associated with a change of ionization and physical conditions across the LIS, shock heating process or photoelectric heating by dust. Intriguingly, the model of the nebula from \cite{Sabbadin2004} shows an increasing $T_{e}$\sulfur~ with the distance from the central star ending with a bump at the position of the LISs (see their fig.~10). As for $N_{e}$, both diagnostics (\sulfurt~and~\chloro) display a very similar radial distribution with clear bumps at 11$<$r$<$15~arcsec and 23$<$r$<$26~arcsec the exact positions of K2/K4 and K1/K4 sub-structures, respectively \citep[see ][]{Goncalves2003}.

Besides the radial distribution of the line fluxes as a function of the distance from the central star, the {\it radial analysis} module also provides the distance from the central star in which each line peaks for a specific region in the nebula defined by the user. For the case of NGC~7009 (Fig.~\ref{figradiallines}), the radial distributions of emission lines are normalized to unity for distances $>$20 arcsec (focused at the NE-LISs). The moderate/high ionization lines (e.g. \oxygeniii, \argoniii) as well as the recombination lines \ha~and \hb~are brighter in the inner nebula and drops outwards from 10 arcsec from the central star and  peak again at the position of the knots (r=23-26 arcsec). On the other hand, the low ionization lines show a first peak at the position of the K2(K3) sub-structures around 10 arcsec and a second one at the position of the K1(K4) knots \citep[][]{Goncalves2003}.

\begin{figure}
\centering
\includegraphics[width=8.65cm]{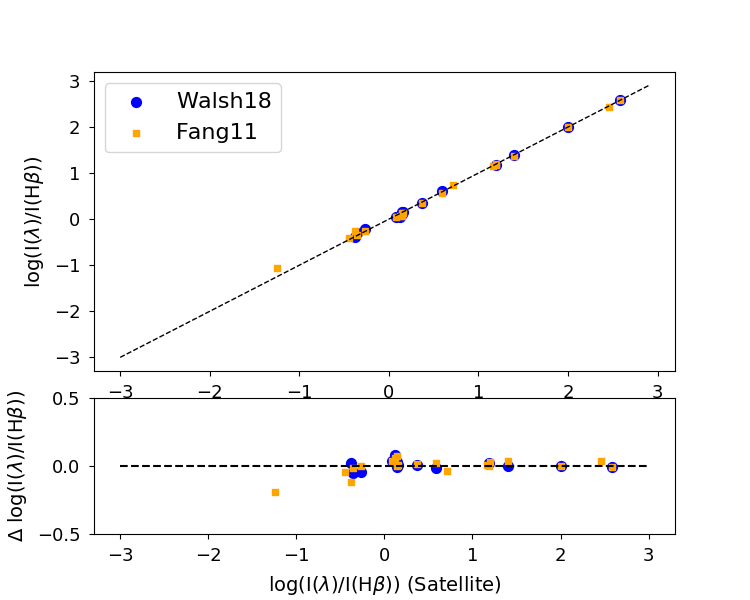}
\caption{Comparison of the line ratios relative to H$\beta$=100 for the pseudo-slit at PA=79~degrees along the major axis of NGC~7009 studied by \protect\cite{Fang2011}. The line ratios from \protect\cite{Fang2011} (labelled as Fang~11) and those reported by \protect\citep[][labelled as Walsh~18]{Walsh2018} are compared to the line ratios from the MUSE data using SATELLITE.}
\label{figlinecomp}
\end{figure}

Note that the moderate/high and low-ionization lines peak at difference radial distances from the central star (see lower panel of Fig.~\ref{figradiallines}). In Table~\ref{distancepeak}, we list the distance from the central star where each line shows an augmentation at the position of $\sim$ 25~arcsec \citep[K1 knot, ][]{Goncalves2003} and a clear stratification is found. In particular, the low-ionization lines (e.g., \nitrogen, \oxygenii, \sulfurt, etc.) peak at a distance of 24.8~arcsec, while the moderate/high ionization lines (e.g., \oxygeniii, \sulfur, \chloro, etc.) display a peak closer to the central star by 0.6-1~arcsec (Fig.~\ref{figradiallines}).

\subsection{Specific slits analysis module}
For a direct comparison of the pseudo-slits spectra with previous long-slit spectroscopic data of NGC~7009 \citep{Goncalves2003,Fang2011}, we employed the {\it specific slits} module to replicate the observations. Line intensities are computed by {\sc satellite} for a number of slit positions/regions (Fig.~\ref{figregions}, Table~\ref{fluxlinesPN}). Fig.~\ref{figlinecomp} displays the comparison between the line intensities computed from {\sc satellite} and those from \cite{Walsh2018} and \cite{Fang2011} for the slit~1. For the interstellar extinction, we consider the same laws used in the studies above i.e., \citet{Howarth1983} for the pseudo-slit~1 and \citet{Cardelli1989} for the rest of the pseudo-slits. It should also be noted that for the estimation of the ionic and total abundances in this module, $T_{e}$\nitrogen~ and $N_{e}$\sulfurt~were considered for the low-ionization species while $T_{e}$\sulfur~ and $N_{e}$\chloro~ were used for the moderate/high ionization species.

A reasonable agreement within the uncertainties is found for most of the line intensities as well as for the physical parameters c(\hb), $T_{e}$, $N_{e}$ and abundances (Table~\ref{fluxlinesPN}). The uncertainties of the physical parameters are calculated considering a Monte Carlo approach and vary from 10 to 30 percent. Higher uncertainties are found for the sub-structures (K or J) because of the small size of the extracted windows (i.e. number of spaxels).

\begin{figure}
\centering
\includegraphics[width=8.5cm]{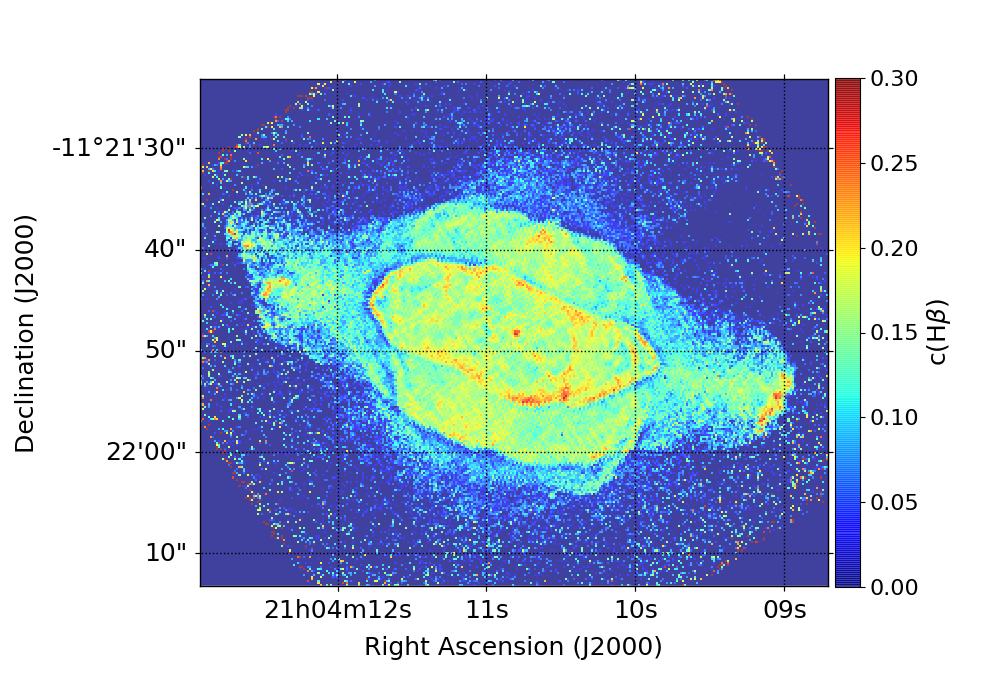}
\includegraphics[width=8.5cm]{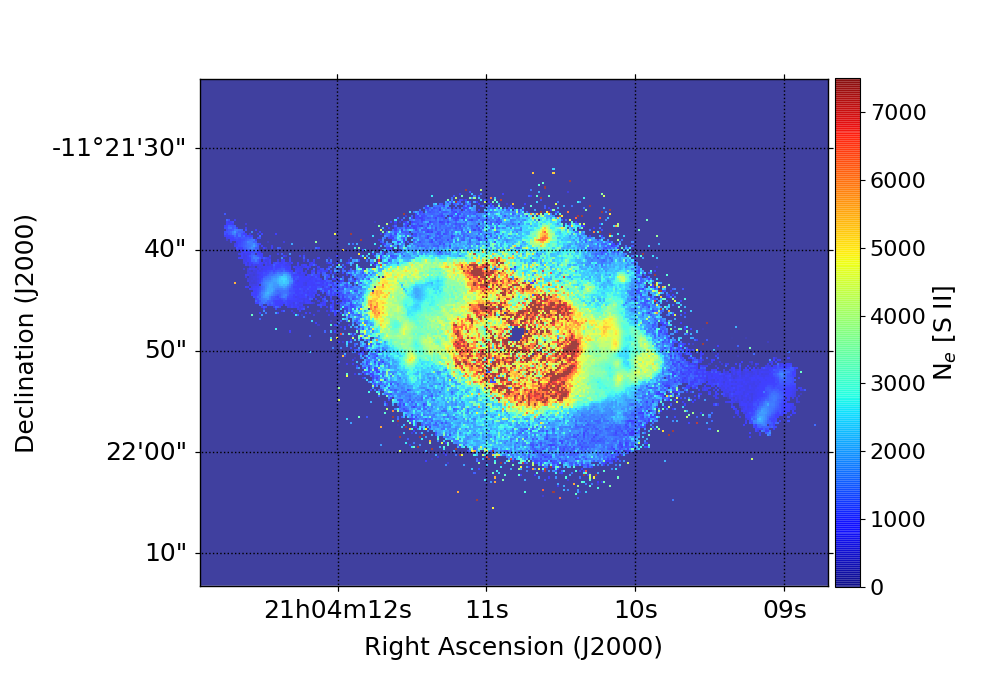}
\includegraphics[width=8.5cm]{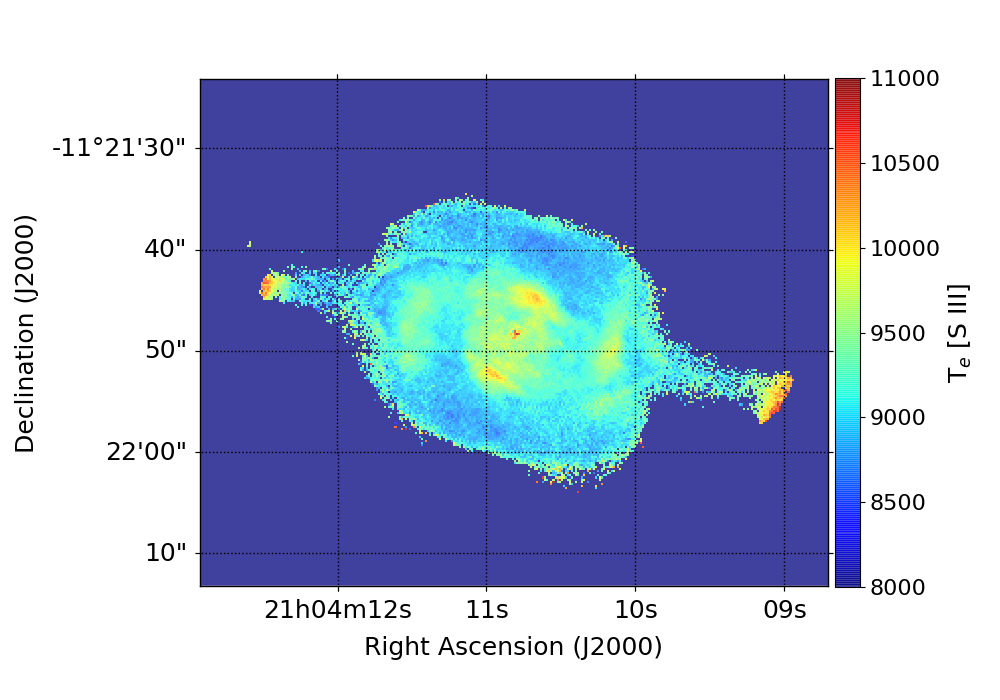}
\caption{Illustrative output maps of the {\sc satellite} code: c(\hb) (upper), $N_{e}$ (middle) and $T_{e}$ (lower panel).}
\label{figchb2Dimage1}
\end{figure}

\begin{figure}
\centering
\includegraphics[width=8.5cm]{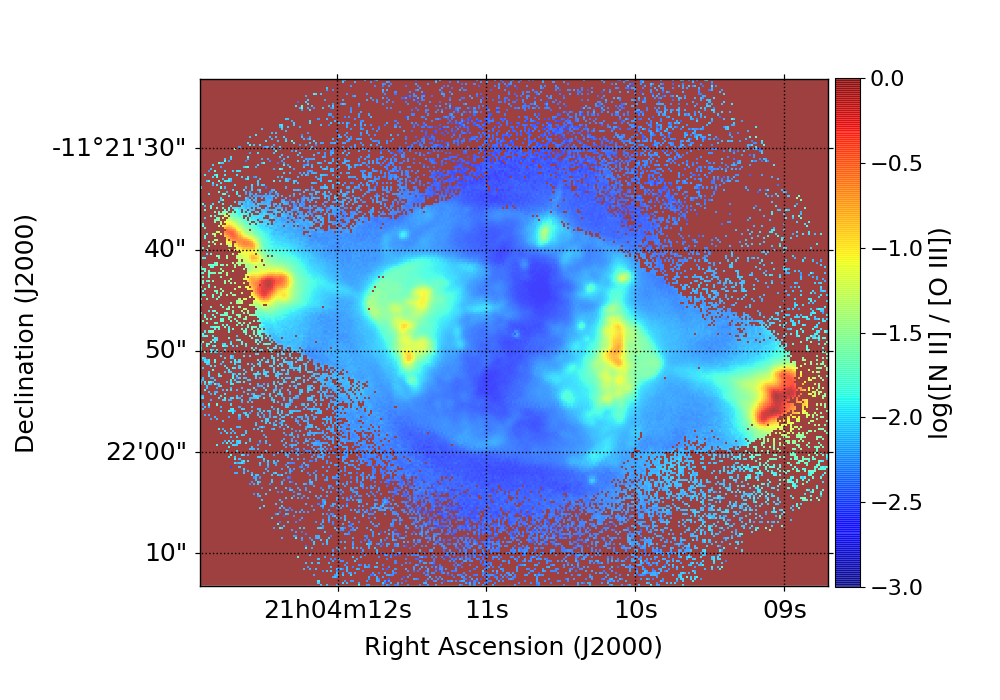}
\includegraphics[width=8.5cm]{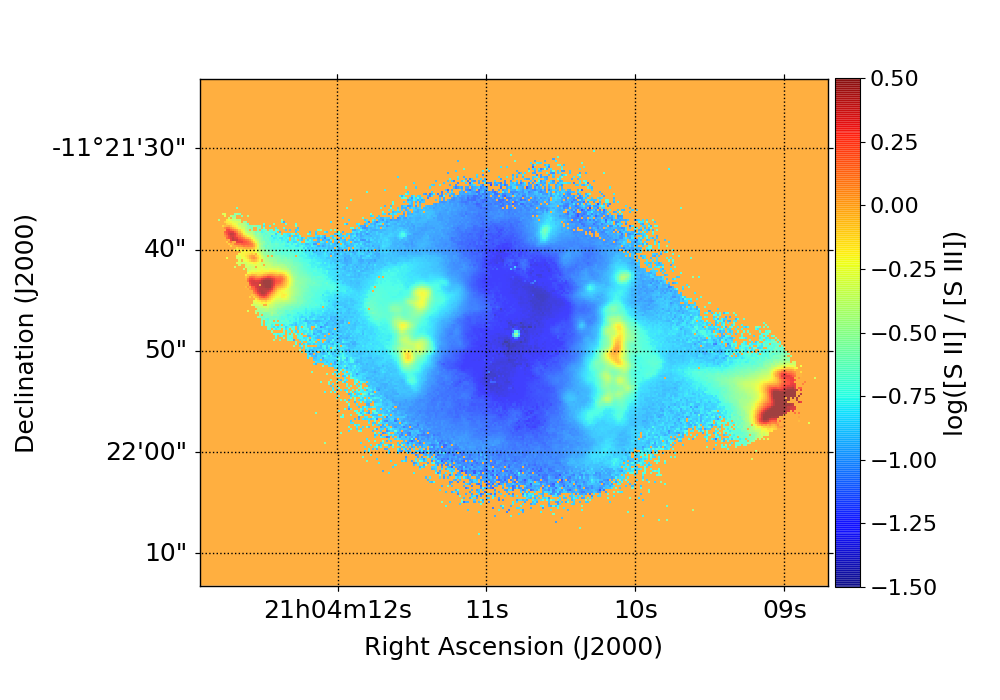}
\caption{Illustrative output maps of the {\sc satellite} code: log(\nitrogen(6548+6584)/\oxygeniii(4959+5007)) (upper) and log(\sulfurt(6716+6731)/\sulfur(6312+9069)) (lower panel) line ratios.}
\label{figchb2Dimage2}
\end{figure}

An analysis on the $T_{e}$ and $N_{e}$ between the different sub-structures and the entire nebula confirms the systematical lower electronic densities found in LISs compared to the main nebula and comparable electronic temperatures \citep[e.g., ][]{Goncalves2003,Goncalves2009,Akras2016}. The spectroscopic analysis of the specific regions/sub-structures shows a reasonable agreement with previous studies and verifies the performance of the {\sc satellite} code. The larger discrepancy is found in $N_{e}$\chloro~for the pseudo-slit~10. We argue that the value of 1300~cm$^{-3}$ \cite[][]{Goncalves2003} is a typo as the \chloro~$\lambda$5517/$\lambda$5538 line ratio is very close to the value computed by {\sc satellite}. Differences in abundances are mainly related to the $T_{e}$/$N_{e}$ diagnostics and the atomic data.

\subsection{2D analysis module}
Regarding the {\it 2D analysis} module, it allows us to study the entire nebula in both spatial directions simultaneously and explore the distribution of the physical parameters. In Figs~\ref{figchb2Dimage1} and \ref{figchb2Dimage2}, we present as illustrative examples the maps of c(\hb), $T_{e}$, and $N_{e}$ as well as the log(\nitrogen/\oxygeniii) and log(\sulfurt/\sulfur) line ratios maps selected from a list of available line ratios. c(\hb) map displays a filamentary structure in the inner nebula with enhanced extinction at the knots \citep[see also fig.~1 in ][]{Walsh2016}. It is evident that c(\hb) is not smooth throughout the nebula and varies from 0.1 up to 0.3. 

\begin{table*}
\caption{Optical emission line fluxes from specific pseudo-slits or nebular regions.} \label{fluxlinesPN}
\begin{tabular}{lcccccccccccc}
\hline

Ion              &I$_{S_F}$ &I$_{W_F)}$& I$_{F}$ & I$_{S_{K1}}$ & I$_{G_{K1}}$ & I$_{S_{K4}}$ & I$_{G_{K4}}$ & I$_{S_{R1}}$ & I$_{G_{R1}}$ & I$_{S_{Neb}}$& I$_{G_{Neb}}$\\
              & p-slit~1 & p-slit~1 & slit 1 & reg.~2 & reg.~2 & reg.~9 & reg.~9 & reg.~5 & reg.~5 & p-slit~10 & slit~10\\
\hline
He~{\sc ii}~4686~\AA\ & -      &  -     &  -      & -      & 0.99 & -     & 1.21 & -     & 25.6   & -     & 15.6 &\\
H~{\sc i}~4861~\AA\   & 100	   & 100    & 100     & 100    & 100  & 100   & 100  & 100   & 100    & 100   & 100  &\\ 
\oiiia                & 385    & 388    & 388     & 395    & 428  & 429   & 454  & 380   & 405    & 390   & 421  &\\ 
\nia                  &  0.048 &  -     & 0.089	  & 4.11   & 6.06 & 4.10  & 2.82 & 0.005 &   -    & 0.143 & 0.11 &\\
He~{\sc ii}~5412~\AA\ &  1.32  &  1.1   & 1.16    & 0.008  & -    & 0.003 &  -   & 1.94  &  1.48  & 1.19  & 1.12 &\\
\cliiia               &  0.435 &  0.5   & 0.454   & 0.860  & -    & 0.932 & 1.03 & 0.377 &  0.46  & 0.566 & 0.58 &\\
\cliiib               &  0.539 &  0.6   & 0.546   & 0.796  & -    & 0.797 & 0.96 & 0.476 &  0.57  & 0.674 & 0.68 &\\
\niia                 &  0.336 &  -     & 0.395   & 5.21   & 6.95 & 5.38  & 4.19 & 0.187 &  0.15  & 0.530 & 0.5  &\\
He~{\sc i}~5876~\AA\  & 14.6   &  -     & 14.4    & 16.4   & 20.3 & 16.2  & 16.2 & 13.8  &  15.0  & 14.8& 15.7 &\\
\oi                   &  0.355 &  0.4   & 0.553   & 25.2   & 31.0 & 21.7  & 14.8 & 0.013 &  -     & 1.01  & 1.06 &\\
\siiia                &  1.39  &  1.4   & 1.39    & 3.07   & 4.29 & 3.39  & 3.59 & 1.19  &  1.41  & 1.73  & 1.86 &\\
H~{\sc i}~6563~\AA\   & 288    &  -     & 266     & 290    & 394  & 290   & 224  & 291   &  312   & 291   & 313 &\\
\niib                 & 14.3   & 14.7   & 15.6    & 324    & 397  & 330   & 217  & 5.05  &  7.93  & 26.5  & 30.2 &\\
He~{\sc i}~6678~\AA\  &  3.88  &  4.0   &  3.68   & 4.4    & 8.41 & 4.40  & 3.57 & 3.71  &  4.35  & 3.93  & 4.46 &\\
\siia                 &  1.28  &  1.4   &  1.38   & 31.4   & 41.4 & 36.8  & 26.1 & 0.458 &  0.54  & 2.50  & 2.63 &\\ 
\siib                 &  2.18  &  2.3   &  2.28   & 43.8   & 57.4 & 45.3  & 32.0 & 0.818 &  0.95  & 4.12  & 4.34 &\\
\ariii                & 15.7   &  -	    & 14.9    & 27.4   & -    & 29.2  &  -   & 14.5  &   -    & 17.5  &  -   &\\
\oiia                 &  1.36  & 1.3:   &  1.19   & 8.56   & -    & 7.45  &  -   & 1.09  &   -    & 1.63  &  -  &\\
\oiib                 &  1.17  & 1.1    &  1.11   & 7.19   & -    & 6.36  &  -   & 0.943 &   -    & 1.38  &  -  &\\
\siiib                & 24.5   & 25.2   & 23.2    & 48.7   & -    & 51.3  &  -   & 19.7  &   -    & 29.9  &  -  &\\
\hline
F(\hb)(10$^{-13}$)    & 199     &-      &  -      & 1.21   & 1.05   & 0.78   & 1.28    & 22.6  &   30.7  & 158    & 188 &\\
c(\hb)                & 0.177   &  -    & 0.174   & 0.158  & 0.16   & 0.16   & 0.16    & 0.165 &   0.16  & 0.165  &  0.16 & \\
Te[N~{\sc ii}]        & 11724   &  -    & 10780   & 10191  & 11000  & 10309  & 11700   & 14735 &  10400  & 10989  & 10300 \\
Te[S~{\sc iii}]       &  9272   &  -    & 11500   & 9677   &9600$^b$&  9877  &10400$^b$&  9453 &10000$^b$&  9293  &10100$^b$  \\
Ne[S~{\sc ii}]        &  3661   &  -    & 4100    & 1628   &  2000  &  1059  &  1300   &  4959 &   5500  &  3100  & 4000 \\
Ne[Cl~{\sc iii}]      &  5150   &  -    & 3600    & 1862   &   -    &  1249  &  1900   &  5475 &   5200  &  4557  & 1300 \\
\hline        
\ch{He+}(5876)/\ch{H+}    & 0.093   &  -    & 0.103    & 0.11       & 0.10   & 0.11      & 0.096&   0.074    & 0.095  & 0.096 & 0.098\\
\ch{He+}(6678)/\ch{H+}    & 0.088   &  -    & 0.095    & 0.11       &  -     & 0.11      &  -   &   0.070    & -      & 0.091 & - \\
\ch{He^{++}}(5412)/\ch{H+}& 0.014   &  -    & 0.013$^c$& 0.00008    &  -     & 0.00003   &  -   &   0.021    & 0.012$^c$& 0.013 & 0.013$^c$\\
He/H                      & 0.109   &  -    & 0.112    & 0.11       & 0.10   & 0.11      & 0.096&   0.094    & 0.108  & 0.106     & 0.111\\
N$^0$/\ch{H+}~(-7)        & 1.13    & -     & 0.84     & 96.8       & 87.0   & 75.3      & 27.0  &  0.07     &  -    & 3.72      & 2.7 \\
\ch{N+}/\ch{H+}~(-6)      & 2.02    & -     & 2.73     & 62.7       & 50.0   & 61.6      & 23.8  &  43.7     & 1.10  & 4.31      & 4.45  \\
ICF(N)$^{a}$              & 67.3/-  & -     & -        & 3.94/7.06  &  7.8   & 4.13/8.44 & 10.6  &  264/-    & 64.7  & 38.0/-    & 38.2\\
N/H$^{a}$~(-5)            & 13.6/-  &-      & 7.01$^d$ & 24.7/44.3  & 38.0   & 25.5/52.1 & 25.0  &  11.6/-   & 7.0   & 16.5/-    & 17.0\\
O$^0$/\ch{H+}~(-6)        & 0.39    & -     & 0.84     & 44.4       & 45.0   & 36.6      & 18.0  &  0.07     & 0.5   & 1.36       & 1.71\\
\ch{O+}/\ch{H+}~(-5)      & 0.87    & -     & 1.99     & 15.6       & 7.5    & 14.9      & 4.2   &  21.5     & 0.7   & 1.53       & 1.2\\
\ch{O^{++}}/\ch{H+}~(-4)  & 5.20    & -     & 3.23     & 4.59       & 5.12   & 4.65      & 4.1   &  4.80     & 4.12  & 5.22       & 4.21 \\
ICF(O)$^{a}$              & 1.10/1.09& -    & 1.086    & 1.0/1.0    & 1.0    & 1.09/1.08 & 1.00  &  1.18/1.16& 1.08  & 1.09/1.08  & 1.08 \\
O/H$^{a}$~(-4)            & 5.83/5.75& -    & 3.6      & 6.16/6.16  & 5.8    & 5.86/5.79 & 4.5   &  5.70/5.59& 4.5   & 5.86/5.79  & 4.71 \\
\ch{S+}/\ch{H+}~(-7)      & 1.06     & -    & 1.19     & 24.0       & 22.1   & 23.2      & 10.1  &  0.28     & 0.44  & 2.17       & 2.09 \\
\ch{S^{++}}/\ch{H+}~(-6)  & 4.18     & -    & 2.26     & 7.74       & 7.45   & 7.86      & 4.8   &  3.24     & 2.1   & 5.11       & 3.3 \\
ICF(S)$^{a}$              & 2.83/-   & -    & ?        & 1.20/1.09  & 1.43   & 1.13/1.11 & 1.57  &  4.46/-   & 2.79  & 2.36/-     & 2.35 \\
S/H$^{a}$~(-6)            & 12.2/-   & -    & 13.0     & 12.1/11.1  & 13.9   & 12.3/11.3 & 9.3   &  14.6/-   & 6.1   & 12.5/-     & 8.3\\
\ch{Cl^{++}}/\ch{H+}~(-8) & 8.65     &  -   & 5.51     & 11.8       &   -    & 11.3      & -     &  7.15      &    - & 10.8       &   -   \\
ICF(Cl)$^{a}$             & -/2.92$^e$& -   & ?        & -/-        &   -    & -/-       & -     &  -/3.90$^e$&    - & -/2.37     &   -   \\
Cl/H$^{a}$~(-7)           & -/2.36$^e$& -   & 1.93     & -/-        &   -    & -/-       & -     &  -/2.79$^e$&    - & -/2.55     &   -   \\
\ch{Ar^{++}}/\ch{H+}~(-6) & 1.56     &  -   & 1.03     & 2.43       &   -    & 2.46      & -     &  1.36      &    - & 1.71       &   -   \\
ICF(Ar)$^{a}$             & 1.87/-   &  -   & ?        & 1.87/1.19  &   -    & 1.87/1.21 & -     &  1.87/-    &    - & 1.87/-     &   -   \\
Ar/H$^{a}$~(-6)           & 2.92/-   &  -   & 2.57     & 4.55/2.91  &  -     & 4.61/2.98 & -     &  2.54/-    &    - & 3.20/-     &   -   \\
\hline
\end{tabular}
\begin{flushleft}
I is the intensity of the lines in the scale where H$\beta$ = 100. The The indices S, W, F, and G point out the results obtained with {\sc satellite} or from previous studies: \citep{Walsh2018}, \citep{Fang2011} and \citep{Goncalves2003}. The indices K1, K4, R and Neb correspond to the sub-structures and total nebula based on \cite{Goncalves2003}. 
$^a$ Elemental abundances are provided using the ICFs from \cite{Kingsburgh1994K} (left) and \cite{Delgado2014} (right), $^b$ These values correspond to the Te[O~{\sc iii}] values, $^c$\heliumb~4686\AA\ line is used for the computation of the ionic abundance, $^d$ Recombination and collisionally excited lines from UV, optical and IR were used to determine the abundance, $^e$ {\sc satellite} does not calculate/provide the total abundance and ICF of Cl because it lies outside the range of validity. \lq\lq ?\rq\rq = unknown value.
\end{flushleft}
\end{table*}

\begin{figure*}
\centering
\includegraphics[width=13.1cm]{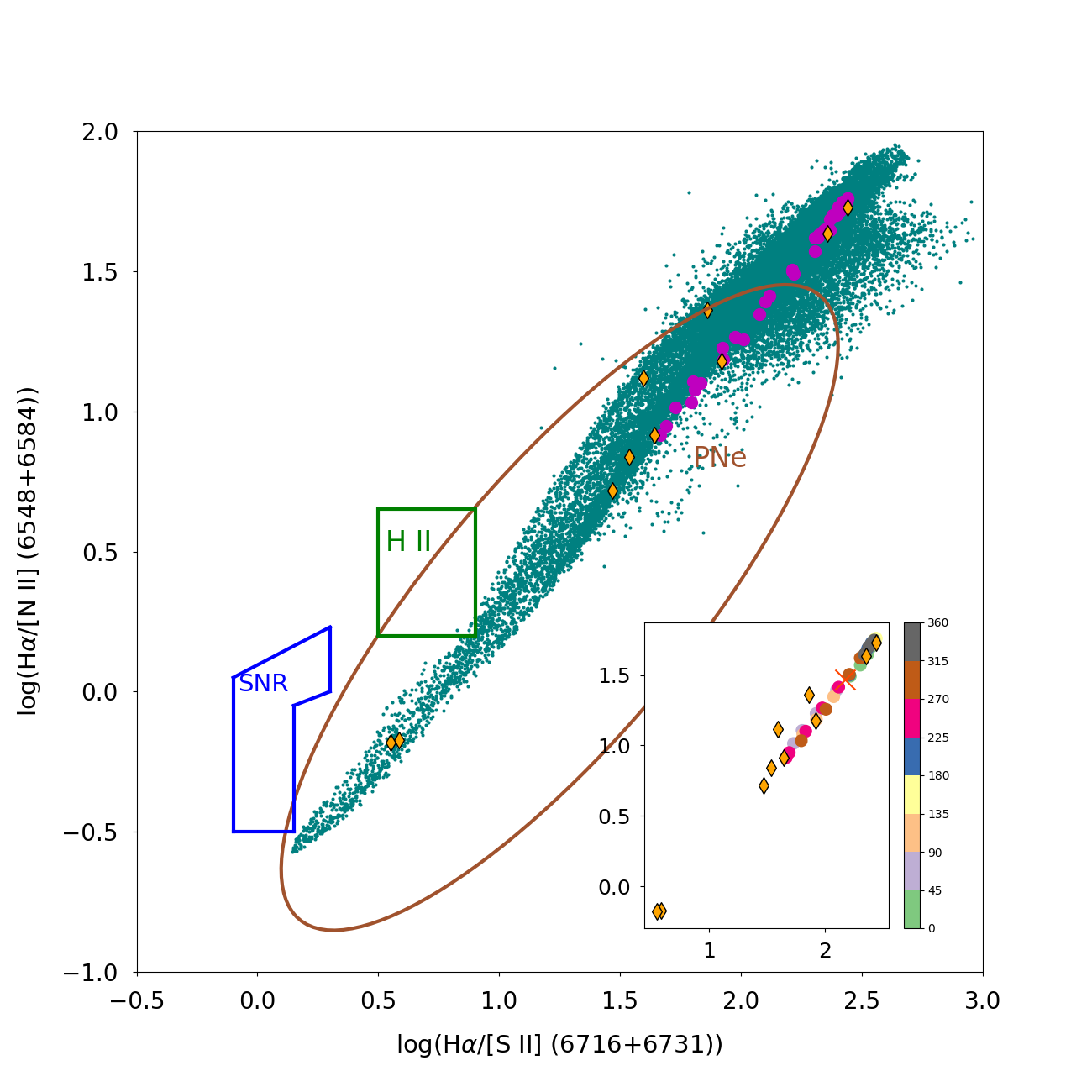}
\caption{Log(\ha/\sulfurt~6716+6731) versus Log(\ha/\nitrogen~6548+6584) diagnostic diagram of NGC~7009 as a representative example. Cyan dots correspond to the values of individual spaxels, pink circles and yellow diamonds show the values obtained from the pseudo-slits of the {\it rotational analysis} module with position angles from 0 to 360 degrees with 10 degrees step and the values from the 10 pseudo-slits in the {\it specific slits analysis} module, respectively. The inset plot illustrate the variation of the line ratios with position angles of the pseudo-slits. The regimes of the PNe, H~{\sc ii} regions and supernova remnants are also drawn \protect\citep{Sabbadin1977,Riesgo2006,Frew2010,Sabin2013}.}
\label{figHaSIIHaNII_DD}
\end{figure*}

Besides the maps, the {\it 2D analysis} module also provides emission line diagnostic diagrams, selected by the user from a predefined list. Fig.~\ref{figHaSIIHaNII_DD} illustrates the common diagnostic diagram log(\ha/\sulfurt) versus log(\ha/\nitrogen). The line ratio values from the individual spaxels are plotted as cyan points, the pseudo-slit spectra from the {\it rotation analysis} module as purple circles in the main plot or colored circles in the inset plot (color bar corresponds to the PA of the pseudo-slits) and those from the {\it specific slits analysis} module as yellow diamonds.

\begin{table*}
\caption{Statistical results of NGC~7009 MUSE datacube.} \label{statMUSEdatacube7009}

\begin{tabular}{llrrrrrrr}

\hline

Parameter   & Num.~of~pix. & 5\% value &  25\% value &  50\% & 75\% value &  95\% value & mean & SD  \\
\hline
c(\hb)                                   & 40841  & 0.011    & 0.053     &  0.108    & 0.145    & 0.193    & 0.105    & 0.065\\
Te(NII6548\_84)\_Ne(SII6716\_31)         & 19554  & 10000    & 10900     &  12100    & 13500    & 18700    & 13000    & 3200\\
Ne(SII6716\_31)\_Te(NII6548\_84)         & 19554  & 828      & 1550      &  2490     & 3660     & 6870     & 3040     & 2090\\
Te(SIII6312\_9069)\_Ne(SII6716\_31)      & 26658  & 8770     & 9000      &  9220     & 9480     & 10800    & 9420     & 749\\
Ne(SII6716\_31)\_Te(SIII6312\_9069)      & 26658  & 615      & 1160      &  1930     & 3030     & 6100     & 2570     & 2300\\
Te(SIII6312\_9069)\_Ne(ClIII5517\_38)    & 17305  & 8780     & 8950      &  9130     & 9270     & 9700     & 9170     & 301\\
Ne(ClIII5517\_38)\_Te(SIII6312\_9069)    & 17305  & 1010     & 2060      &  3220     & 4610     & 7070     & 3610     & 2040\\
\hline                                     
log(He~{\sc i}~5876/\ha)                                &  40017 & -1.33  & -1.27  &  -1.26  & -1.25  & -1.22  & -1.26  & 0.04\\
log(He~{\sc ii}~5412/\hb)                               &  14528 & -3.05  & -2.88  &  -2.53  & -2.13  & -1.64  & -2.44  & 0.48\\
log(He~{\sc i}~5876/He~{\sc ii}~5412)                   &  14526 &  0.76  & 1.22   &   1.74  &  2.05  &  2.26  &  1.63  & 0.51\\
log(\nitrogen~6584/\ha)                                 &  40228 & -1.87  & -1.70  &  -1.55  & -1.42  & -0.69  & -1.46  & 0.38\\
log((\nitrogen~6548,6584)/(\oxygeniii~4959,5007))       &  40214 & -2.52  & -2.39  &  -2.25  & -2.11  & -1.35  & -2.15  & 0.38\\
log(\nitrogena~5200/\hb)                                &   5232 & -3.22  & -3.06  &  -2.64  & -2.23  & -1.10  & -2.48  & 0.66\\
log((\sulfurt~6716,6731)/\ha)                           &  28808 & -2.46  & -2.25  &  -2.08  & -1.90  & -1.14  & -1.98  & 0.42\\
log(\sulfurt~6716/\sulfurt~6731)                        &  28808 & -0.29  & -0.23  &  -0.17  & -0.11  & 0.0001 & -0.16 & 0.09\\
log((\sulfurt~6716,6731)/(\sulfur~6312,9069))           &  27687 & -1.27  & -1.06  &  -0.91  & -0.82  & -0.32  & -0.88 & 0.30\\
log\oxygeni~6300/\ha)                                   &  11752 & -3.80  & -3.71  &  -3.16  & -2.62  & -1.25  & -2.95  & 0.83\\
log(\oxygeniii~5007/\hb)                                &  40734 &  1.05  & 1.07   &   1.12  &  1.20  &  1.28  &  1.14  & 0.08\\
log((\oxygenii~7320,7330)/(\oxygeniii~4959,5007))       &  22637 & -3.15  & -3.06  &  -2.97  & -2.86  & -2.45  & -2.91  & 0.24\\
log((\chloro~5517,5538)/\hb)                            &  18176 & -2.10  & -2.03  &  -1.97  & -1.92  & -1.80  & -1.96  & 0.09\\
\hline
abundance~(He~{\sc i}~5876\AA)                 & 17305 & 0.088    & 0.102    & 0.108   &  0.110    & 0.113    & 0.105    & 0.008\\
abundance~(He~{\sc i}~6678\AA)                 & 17305 & 0.083    & 0.096    & 0.103    & 0.105    & 0.109    & 0.099    & 0.008\\
abundance~(He~{\sc ii}~5412\AA)                & 14052 & 9.39e-04 & 1.39e-03 & 3.09e-03 & 8.16e-03 & 2.48e-02 & 7.21e-03 & 8.13e-03\\
abundance~(\oi)                                &  8948 & 1.14e-07 & 1.47e-07 & 2.83e-07 & 7.71e-07 & 1.75e-05 & 4.32e-06 & 1.61e-05\\
abundance~(\oiia)                              & 17298 & 1.69e-05 & 2.40e-05 & 2.90e-05 & 3.30e-05 & 6.60e-05 & 3.49e-05 & 3.15e-05\\
abundance~(\oiib)                              & 17285 & 1.82e-05 & 2.59e-05 & 3.14e-05 & 3.56e-05 & 6.48e-05 & 3.72e-05 & 3.29e-05\\
abundance~(\oiiib)                             & 17305 & 4.44e-04 & 5.32e-04 & 5.73e-04 & 6.01e-04 & 6.62e-04 & 5.67e-04 & 6.85e-05\\
abundance~(\nia)                               &  3646 & 3.22e-07 & 4.10e-07 & 6.73e-07 & 1.52e-06 & 1.53e-05 & 3.05e-06 & 7.26e-06\\
abundance~(\niia)                              & 16705 & 2.64e-06 & 3.29e-06 & 4.36e-06 & 5.96e-06 & 2.42e-05 & 8.37e-06 & 1.52e-05\\
abundance~(\niib)                              & 17305 & 8.35e-07 & 1.34e-06 & 2.03e-06 & 3.21e-06 & 1.85e-05 & 5.69e-06 & 1.44e-05\\
abundance~(\siia)                              & 17301 & 5.17e-08 & 8.23e-08 & 1.29e-07 & 2.15e-07 & 1.00e-06 & 3.18e-07 & 6.91e-07\\
abundance~(\siib)                              & 17305 & 5.06e-08 & 7.97e-08 & 1.20e-07 & 2.00e-07 & 9.18e-07 & 2.93e-07 & 6.11e-07\\
abundance~(\siiia)                             & 17305 & 2.63e-06 & 3.35e-06 & 4.06e-06 & 5.00e-06 & 7.48e-06 & 4.49e-06 & 1.55e-06\\
abundance~(\siiib)                             & 17305 & 2.63e-06 & 3.35e-06 & 4.06e-06 & 5.00e-06 & 7.48e-06 & 4.49e-06 & 1.55e-06\\
abundance~(\cliiia)                            & 17305 & 6.57e-08 & 8.47e-08 & 9.83e-08 & 1.10e-07 & 1.36e-07 & 9.98e-08 & 2.36e-08\\
abundance~(\cliiib)                            & 17305 & 6.57e-08 & 8.47e-08 & 9.83e-08 & 1.10e-07 & 1.36e-07 & 9.98e-08 & 2.36e-08\\
abundance~(\ariii)                             & 17305 & 1.16e-06 & 1.40e-06 & 1.60e-06 & 1.79e-06 & 2.25e-06 & 1.65e-06 & 3.37e-07\\
\hline
\end{tabular}
\begin{flushleft}
50\% percentiles corresponds to the median.
\end{flushleft}
\end{table*}

The majority of the spaxels as well as some pseudo-slits from the {\it rotation analysis} and {\it specific slit analysis} modules (pseudo-slits~1 and 10 that cover the entire nebula from one side to the other) have values up to log(\ha/\sulfurt)$\sim$2.5 and log(\ha/\nitrogen)$\sim$1.75 lying outside the 85~percent confidence level ellipse \citep{Riesgo2006}. This clearly demonstrates that the regime of PNe in the STB diagnostic diagram should be updated by covering higher \ha/\sulfurt~and \ha/\nitrogen~line ratios. The same conclusion has been reached from the prediction of 1D photo-ionization models \citep[see Fig.~5 in ][]{Akras2020a}. 

Moreover, there is a non-negligible number of spaxels with moderate or even low values (log(\ha/\sulfurt)<1.0 and log(\ha/\nitrogen)<0.25) close to the loci of H~{\sc ii} regions and SNRs \citep{Frew2010,Sabin2013} whereas pseudo-slits from the {\it rotation analysis} and {\it specific slit analysis} modules do not display such low ratios. Interestingly, two pseudo-slits from the {\it specific slit analysis} module (pseudo-slits~2 and 9) have the lowest values (log(\ha/\sulfurt)$\sim$0.5 and log(\ha/\nitrogen)$\sim$ -0.5) and correspond to the extracted windows from the two LISs/knots K1 and K4. Our mean values of the \ha/\sulfurt~and \ha/\nitrogen~ ratios for the K1 and K4 knots agree with the results from HST imaging \cite{Phillips2010} but we do not observe the same distribution for individual spaxels. In particular, HST data display \ha/\sulfurt~and \ha/\nitrogen~ ratio values well distributed in the SNRs regime (see fig.~19 in \citealt{Phillips2010}) while the distribution of the individual MUSE spaxels is within the PNe regime. Hence, we argue that any contribution of shocks in the knots of NGC~7009 is indistinguishable.

It is evident that line ratios can vary significantly from one sub-structure to another. The main nebula of NGC~7009 is certainly UV dominated and all its physical properties as well as emission line ratios can be explained from the strong UV radiation field of its central star. However, there are sub-structures like K1 to K4 LISs that diverge from the main bulk (Fig.~\ref{figHaSIIHaNII_DD}). To explain the large divergence of line ratios in different sub-structures in the nebula with comparable physical parameters (Table~\ref{fluxlinesPN}), a high density gas should be considered in these sub-structures. The detection of H$_2$ emission from the LISs in NGC~7009 \citep{Akras2020b} supports the presence of a high density gas. The enhancement of low-ionisation emission such as \nitrogen, \oxygeni~emission lines in conjunction with the H$_2$ emission implies the presence of mini-photodissociation regions (PDRs). Furthermore, the detection of an emission at 8727\AA\ very likely associated with the red-\carboni~line (Akras et al. 2021 in preparation), found in PDRs \citep[e.g.]{Burton1992} and PNe \citep{Liu1995CI}, further supports the scenario that the enhanced emission from the low-ionization species found in the LISs of NGC~7009 originate from the partially ionized gas of mini-PDRs rather than shock-heated gas.

{\sc satellite} also calculates the mean value, standard deviation and the percentiles of 5\%, 25\% (Q1), 50\% (median), 75\% (Q3), 95\% for each parameter with 2D maps constructed, such as c(\hb), $T_{e}$, and $N_{e}$, ionic abundances and line ratios (Table~\ref{statMUSEdatacube7009}) \citep[see also,][]{Monreal2020}. 

\section{The case study of NGC~7009 with deeper MUSE data}
\label{Cordata}
Besides the SV MUSE data, deeper observations were also obtained in visitor mode on 07 July 2016 (097.D-0241(A), PI: R. L. M. Corradi; hereafter Cor MUSE data) using the extended mode. Short and long observations were obtained with exposure times of 30 and 150~sec, respectively. For the long exposure observations ten frames were obtained at different position angles resulting in total integrated times of 1500~sec. The DIMM seeing during the observations varied from 0.8 to 1.2~arcsec and the airmass changed from 1.126 to 1.357. Unfortunately, these observations were not carried out under photometric conditions due to the presence of thin cirrus. For consistency with the SV data, we found that the short and long exposure line maps of the Cor MUSE datacubes should be multiplied by 0.46$\pm$0.05 and 2.4$\pm$0.1, respectively. This introduces an extra uncertainty on the analysis but it is valuable to verify the performance of {\sc satellite} using the flux maps from a second datacube and an independent reduction and emission line fitting.

\begin{figure}
\centering
\includegraphics[width=8.65cm]{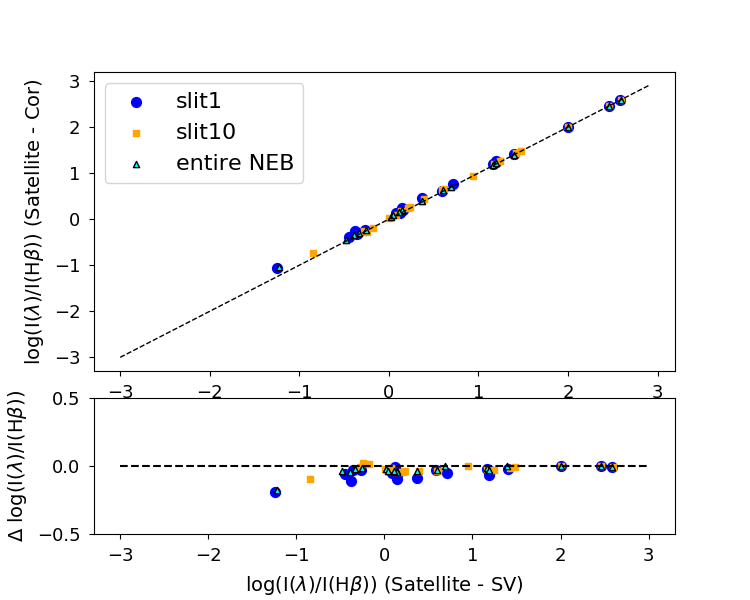}
\caption{Comparison of the line ratios relative to H$\beta$=100 obtained from the pseudo-slits~1, 10 and the entire nebula of NGC~7009 from two MUSE datacubes using {\sc satellite}: science verification MUSE data (SV), \protect\citep{Walsh2018} and the Cor MUSE data (097.D-0241(A), PI: R. L. M. Corradi).}
\label{figlinecomp2}
\end{figure}

Line intensities, physical parameters (c(\hb), T$_{\rm e}$, N$_{\rm e}$) and abundances computed from the pseudo-slits~1, 10 (Fig.~\ref{figregions}) and the entire nebula (2310~square arcseconds) obtained with the {\sc satellite} code from the SV and Cor MUSE data are listed in Table~\ref{fluxlinesComp}. The comparison of the line intensities between the two datacubes is displayed in Fig.~\ref{figlinecomp2}. A reasonable matching is also found for all the nebular parameters. Only the extinction coefficient exhibits an unexpected mismatch of the order or 25~percent, due to the non-photometric night. Thus, within the involved uncertainties, the results validate the {\it specific slit analysis} module of the {\sc satellite} code for the spectroscopic analysis of specific regions or sub-structures in PNe such as knots.

The results from the {\it rotation analysis} and {\it radial analysis} modules are also consistent between the two datasets. All the nebula parameters show the same behaviour with the PA of the pseudo-slits and similar absolute values (not presented here). The high T$_{\rm e}$[N~{\sc ii}] values at the PAs of 0, 150 and 320 degrees are found again with the Cor MUSE data. The radial profiles of the nebular parameter are also comparable. T$_{\rm e}$[N~{\sc ii}] is higher at the central region of the nebula and decreases for larger distance while T$_{\rm e}$[S~{\sc iii}] is nearly constant. As for the electron density, very small differences are found.

For the {\it 2D analysis} module, the results of the statistical analysis of the Cor MUSE maps, for instance the mean values, standard deviations and the 5\%, 25\% (Q1), 50\% (median), 75\% (Q3) and 95\% percentiles provide an overall consistency between the two set datacubes.

\begin{figure}
\centering
\includegraphics[width=7.2cm]{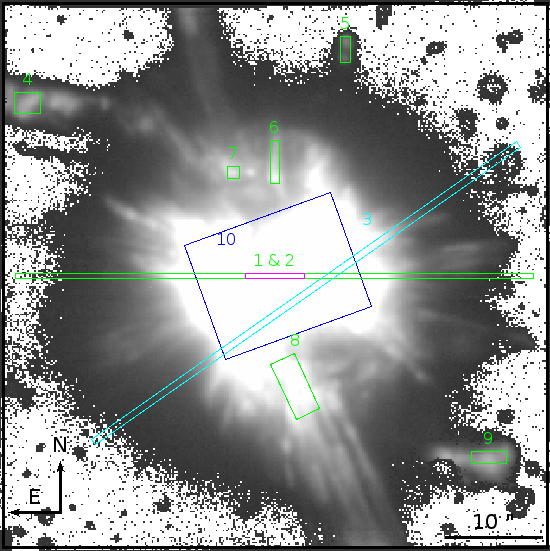}
\caption{10 slits/regions selected for the spectroscopic analysis of specific slits with {\sc satellite} are overlaid on the \nitrogen~$\lambda$6584 line map of NGC~6778. North is up east to the left.} 
\label{figregions6778}
\end{figure} 

\begin{table*}
\caption{Comparison between optical emission line fluxes and derived quantities from specific slits between SV and Cor MUSE data of NGC\,7009.} \label{fluxlinesComp}
\begin{tabular}{lccccccc}
\hline

Ion              &I$_{S_F}$ (SV) & I$_{S_F}$ (Cor) & I$_{S_{Neb}}$ (SV) & I$_{S_{Neb}}$ (Cor) & I$_{S_{entire}}$ (SV) & I$_{S_{entire}}$ (Cor) \\
              & p-slit~1 & p-slit 1 & p-slit~10 & p-slit~10 & 2310~arcsec$^2$ & 2310~arcsec$^2$\\
\hline
He~{\sc ii}~4686~\AA\ & -      &  17.5  & -       & 16.4  & -      & 14.3\\
H~{\sc i}~4861~\AA\   & 100	   &  100   & 100     & 100    & 100   & 100\\ 
\oiiia                & 385    &  391   & 390     & 396    & 388   & 395\\ 
\nia                  &  0.048 &  0.087	& 0.143   & 0.177  & 0.058 & 0.091\\
He~{\sc ii}~5412~\AA\ &  1.32  &  1.35  & 1.20    & 1.24   & 1.04  & 1.10\\
\cliiia               &  0.435 &  0.468 & 0.566   & 0.538  & 0.468 & 0.488\\
\cliiib               &  0.539 &  0.578 & 0.674   & 0.645  & 0.554 & 0.575\\
\niia                 &  0.336 &  0.408 & 0.530   & 0.540  & 0.330 & 0.355\\
He~{\sc i}~5876~\AA\  & 14.6   & 15.4   & 14.8    & 15.6   & 14.9  & 15.7 \\
\oi                   &  0.355 &  0.544	&  1.01   & 1.05   & 0.411 & 0.450\\
\siiia                &  1.39  &  1.60  &  1.73   & 1.86   & 1.39  & 1.50 \\
\niic                 &  4.88  &  5.82  &  8.85   & 8.82   & 4.89  & 4.88 \\
H~{\sc i}~6563~\AA\   & 288    &  289   & 291     & 289    & 288   & 289\\
\niib                 & 14.3   & 18.1   & 26.5    & 27.3   & 14.6  & 15.2 \\
He~{\sc i}~6678~\AA\  &  3.88  &  4.12  &  3.93   & 4.20   & 3.98  & 4.24\\
\siia                 &  1.28  &  1.71  &  2.50   & 2.70   & 1.39  & 1.54 \\ 
\siib                 &  2.18  &  2.88  &  4.12   & 4.41   & 2.32  & 2.53 \\
\ariii                & 15.7   & 17.3   & 17.5    & 18.6   & 15.5  & 16.5 \\
\oiia                 &  1.36  &  1.58  &  1.63   & 1.78   & 1.29  & 1.41 \\
\oiib                 &  1.17  &  1.35  &  1.38   & 1.51   & 1.11  & 1.21 \\
\siiib                & 24.5   & 26.3   & 29.9    & 30.3   & 24.8  & 24.9\\
\hline
F(\hb)(10$^{-13}$)    & 199     &  208    & 158    & 159      & 1458   & 1469\\
c(\hb)                & 0.170   & 0.221   & 0.165  & 0.211    & 0.166  & 0.218\\
Te[N~{\sc ii}]        & 11724   & 11578   & 10989  & 10980    & 11547  & 11822\\
Te[S~{\sc iii}]       &  9272   & 9516    & 9293   & 9566     & 9219   & 9498\\
Ne[S~{\sc ii}]        &  3661   & 3493    & 3100   & 2999     & 3333   & 3167\\
Ne[Cl~{\sc iii}]      &  5150   & 5127    & 4577   & 4776     & 4648   & 4435\\
\hline        
\ch{He+}(5876)/\ch{H+}      & 0.093     & 0.098     &  0.096     & 0.102       & 0.096     &  0.10\\
\ch{He+}(6678)/\ch{H+}      & 0.088     & 0.094     &  0.091     & 0.098       & 0.091     & 0.096\\
\ch{He^{++}}(5412)/\ch{H+}  & 0.014     & 0.014$^b$ &  0.013     & 0.013$^b$   & 0.011     &0.017$^b$\\
He/H                        & 0.109     & 0.111     &  0.106     & 0.113       & 0.105      & 0.107\\
N$^0$/\ch{H+}~(-7)          & 1.13      & 2.05      &  3.72      & 4.50        & 1.34      & 1.88\\
N+/\ch{H+}~(-6)             & 2.02      & 2.57      &  4.31      & 4.41        & 2.12      & 2.06\\
ICF(N)$^{a}$                & 67.3/-    & 40.7/-    &  38.0/-    & 31.4/-      & 64.1/-    & 63.1/-\\
N/H$^{a}$~(-5)              & 13.6/-    & 12.9/-    &  16.5/-    & 13.9/-      & 13.7/-    & 13.8/-\\
\ch{O+}/\ch{H+}~(-5)        & 0.87      & 1.09      &  1.53       & 1.72        & 0.92      & 0.95\\
\ch{O^{++}}/\ch{H+}~(-4)    & 5.20      & 4.82      &  5.22       & 4.79        & 5.35      & 4.90\\
ICF(O)$^{a}$                & 1.10/1.09 & 1.09/1.08 &  1.09/1.08  & 1.09/1.07   & 1.08/1.07 & 1.08/1.07\\
O/H$^{a}$~(-4)              & 5.83/5.75 & 5.41/5.34 &  5.86/5.79  & 5.40/5.33   & 5.87/5.81 & 5.36/5.30\\
\ch{S+}/\ch{H+}~(-7)        & 1.06      & 1.41      &  2.17       & 2.32        & 1.13      & 1.15\\
\ch{S^{++}}/\ch{H+}~(-6)    & 4.18      & 4.27      &  5.11       & 4.86        & 4.30      & 4.07\\
ICF(S)$^{a}$                & 2.83/-    & 2.56/-    &  2.36/-     & 2.21/-      & 2.78/-    & 2.77/-\\
S/H$^{a}$~(-6)              & 12.2/-    & 11.3/-    &  12.5/-     & 11.3/-      & 12.3/-    & 11.6/-\\
\ch{Cl^{++}}/\ch{H+}~(-8)   & 8.65      & 8.55      & 10.8        & 9.41        & 9.10      & 8.60\\
ICF(Cl)$^{a}$               & -/2.92$^c$& -/2.53    & -/2.37      & -/2.26      & -/2.65    & -/2.65\\
Cl/H$^{a}$~(-7)             & -/2.36$^c$& -/2.17    & -/2.55      & -/2.13      & -/2.42    & -/2.23  \\
\ch{Ar^{++}}/\ch{H+}~(-6)   & 1.56      & 1.61      & 1.71        & 1.71        & 1.57      & 1.55\\
ICF(Ar)$^{a}$               & 1.87/-    & 1.87/-    & 1.87/-      & 1.87/-      & 1.87/-    & 1.87/-\\
Ar/H$^{a}$~(-6)             & 2.92/-    & 3.01/-    & 3.20/-      & 3.20/-      & 2.93/-    & 2.89/-\\
\hline
\end{tabular}
\begin{flushleft}
$^{a}$ Elemental abundances are provided using the ICFs from \cite{Kingsburgh1994K} (left) and \cite{Delgado2014} (right). \\
$^b$ \heliumb~4686\AA\ line is used for the computation of the ionic abundance.\\
$^c$ {\sc satellite} does not calculate/provide the total abundance and ICF of Cl because the criteria are not satisfied.\\
\end{flushleft}
\end{table*}

\begin{figure}
\centering
\includegraphics[width=7.8cm]{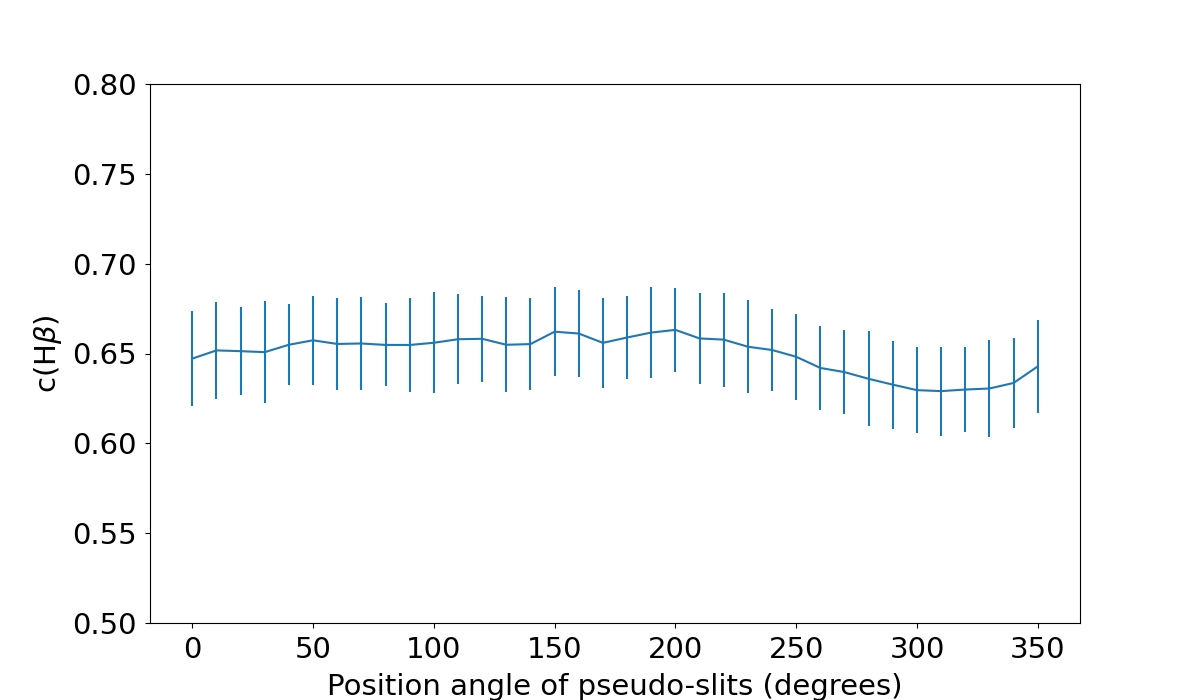}
\includegraphics[width=7.8cm]{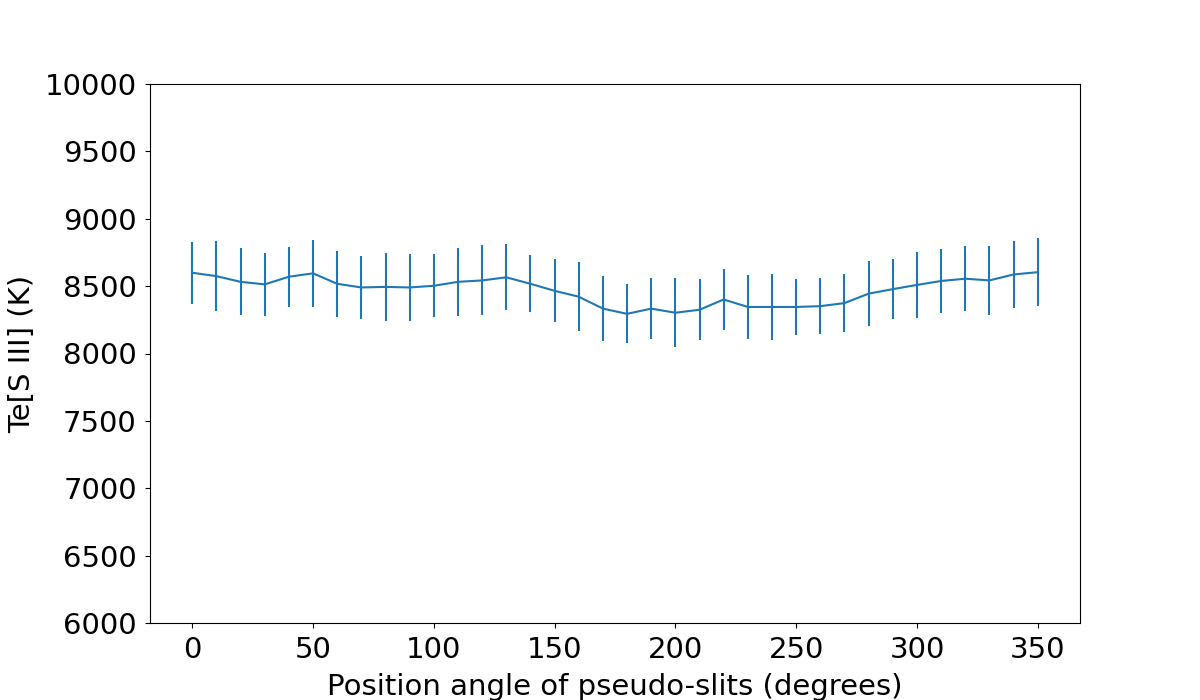}
\includegraphics[width=7.8cm]{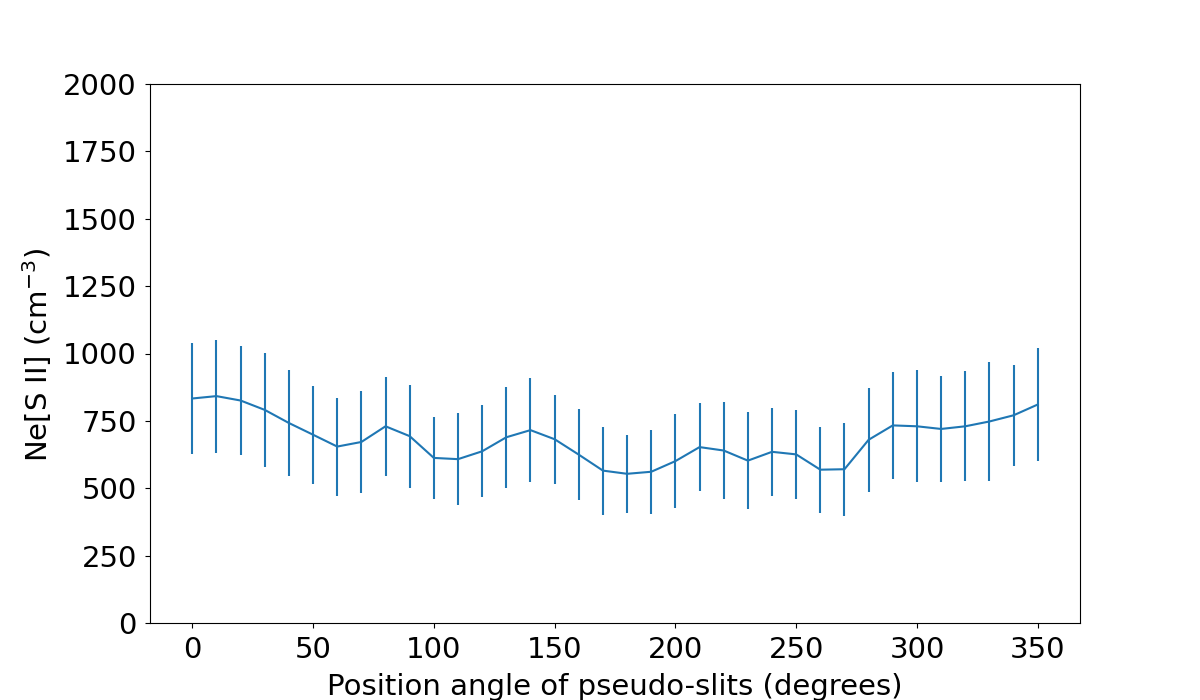}
\caption{c(H$\beta$), T$_{\rm e}$\sulfur~and N$_{\rm e}$\sulfurt~as functions of the pseudo-slit position angles for NGC~6778} 
\label{figTeNePA6778}
\end{figure} 

\section{The case study of NGC~6778 with Cor MUSE data}

In this section, we present the spectroscopic analysis of the planetary nebula NGC~6778 employing the {\sc satellite} code and MUSE data (097.D-0241(A), PI: R. L. M. Corradi). NGC~6778 has a complex morphology with a bright waist, a number of low-ionization knots and two pairs of collimated jets \citep{Guerrero2012}. What makes this nebula an intriguing object is its central star which has been found to be a close binary system with a period of $\sim$0.15~days \citep{Miszalski2011}. Binary systems in PNe have been linked with a high abundance discrepancy factor and NGC~6778 is one of them \citep[adf=20, ][]{Jones2016}.

Recent data from OSIRIS Blue Tunable Filter (GTC) and VIMOS IFU (VLT) have revealed a significant difference in the spatial distribution between the \oxygeniirec~4649+50\AA~ optical recombination lines (ORLs) and the collisionally excited \oxygeniii~$\lambda$5007 line \citep{GarciaRojas2016}. Interestingly, the auroral \oxygeniii~$\lambda$4363 line displays the same spatial distribution as the \oxygeniirec~recombination line resulting in questionable T$_{\rm e}$ \citep{GomezLlanos2020}. These results make NGC~6778 an ideal target for a 2D imaging spectroscopic analysis with deep MUSE data. A full presentation of the data and spatial analysis of the complete MUSE data set of NGC~6778, including \carboniirec, \oxygeniirec~and \nitrogeniirec~ORLs has been presented in \citet{Garciarojas2022}. In this work with {\sc satellite}, we focus on the spectroscopic analysis of the nebula using the bright collisionally excited lines (CELs) and H/He recombination lines.

\begin{figure}
\centering
\includegraphics[width=7.8cm]{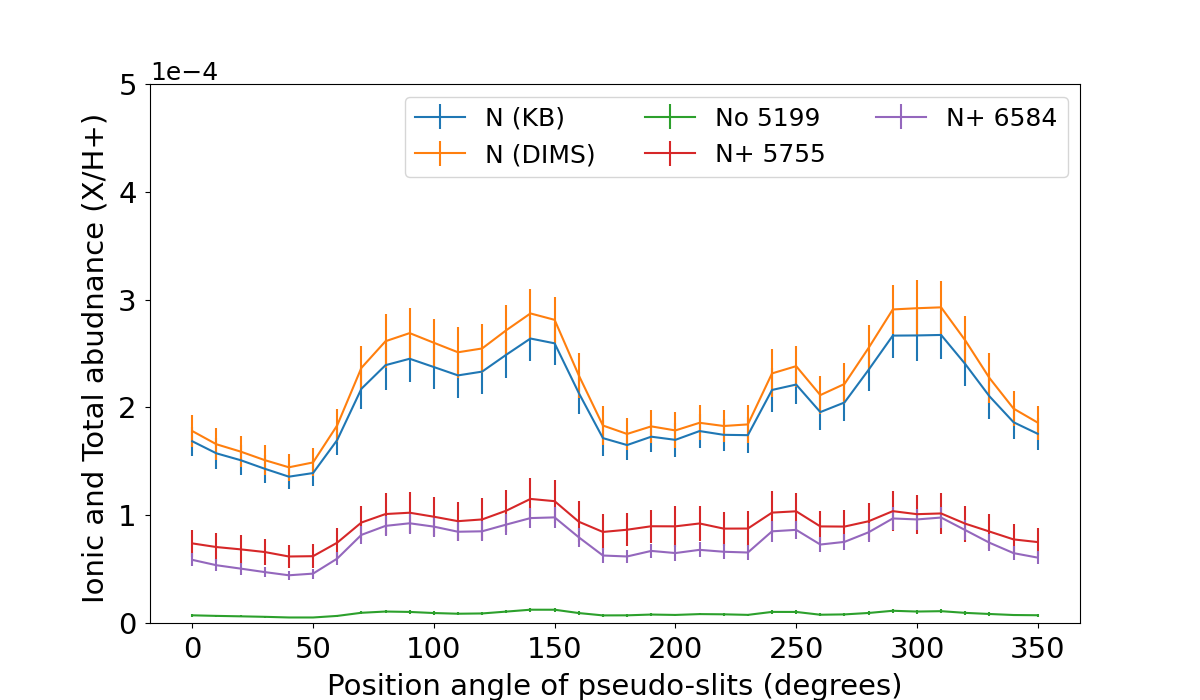}
\includegraphics[width=7.8cm]{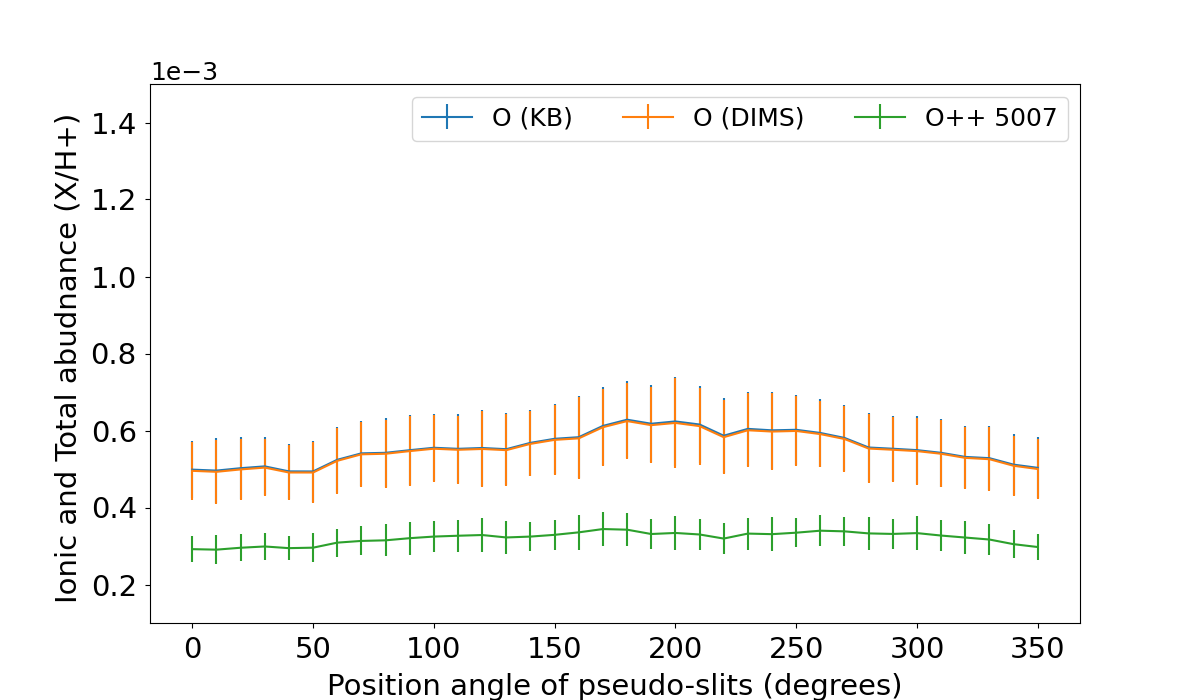}
\includegraphics[width=7.8cm]{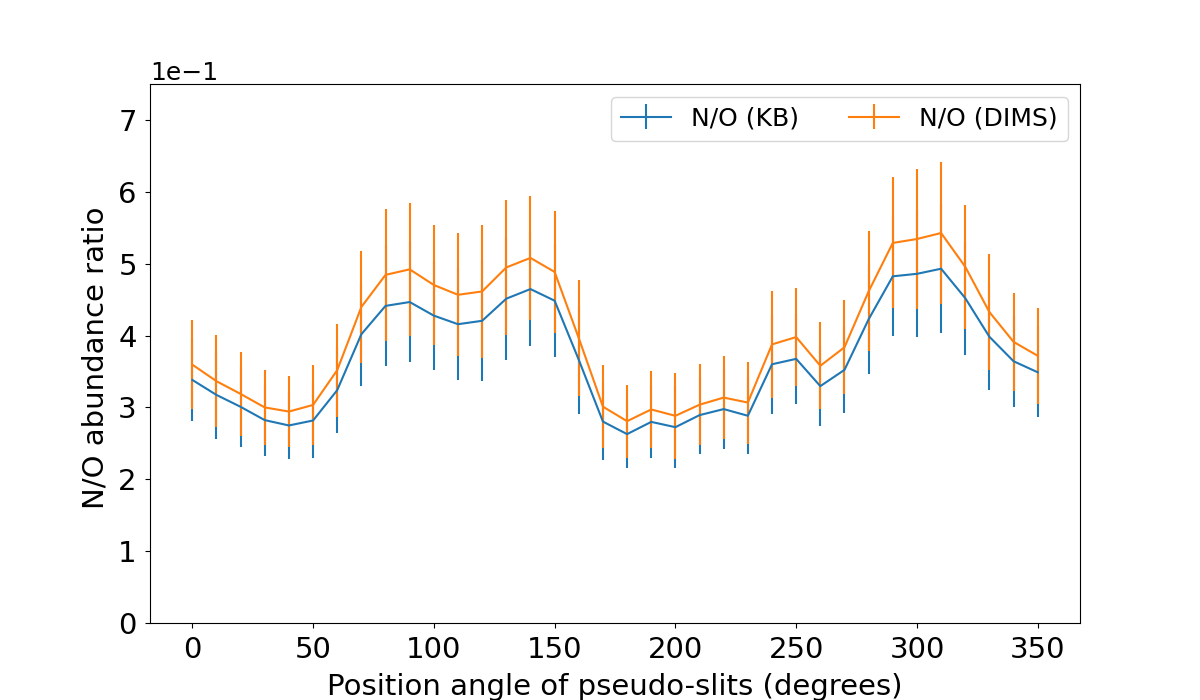}
\caption{Ionic and total N (upper panel) and O (middle panel) abundances of NGC~6778 as well a N/O ratio (lower panel) as a function of the pseudo-slit position angles. The total abundances is computed using the ICFs schemes from \protect\cite{Kingsburgh1994K} and \protect\cite{Delgado2014}.} 
\label{figONabundPA6778}
\end{figure} 

\begin{figure}
\centering
\includegraphics[width=7.8cm]{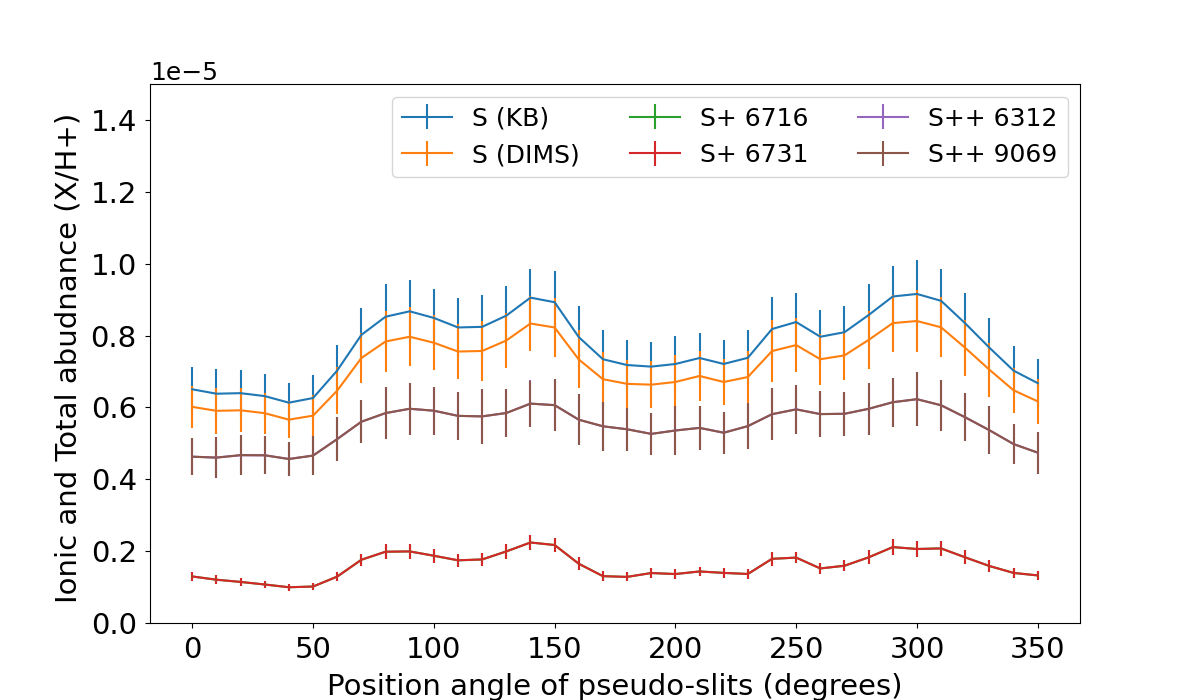}
\includegraphics[width=7.8cm]{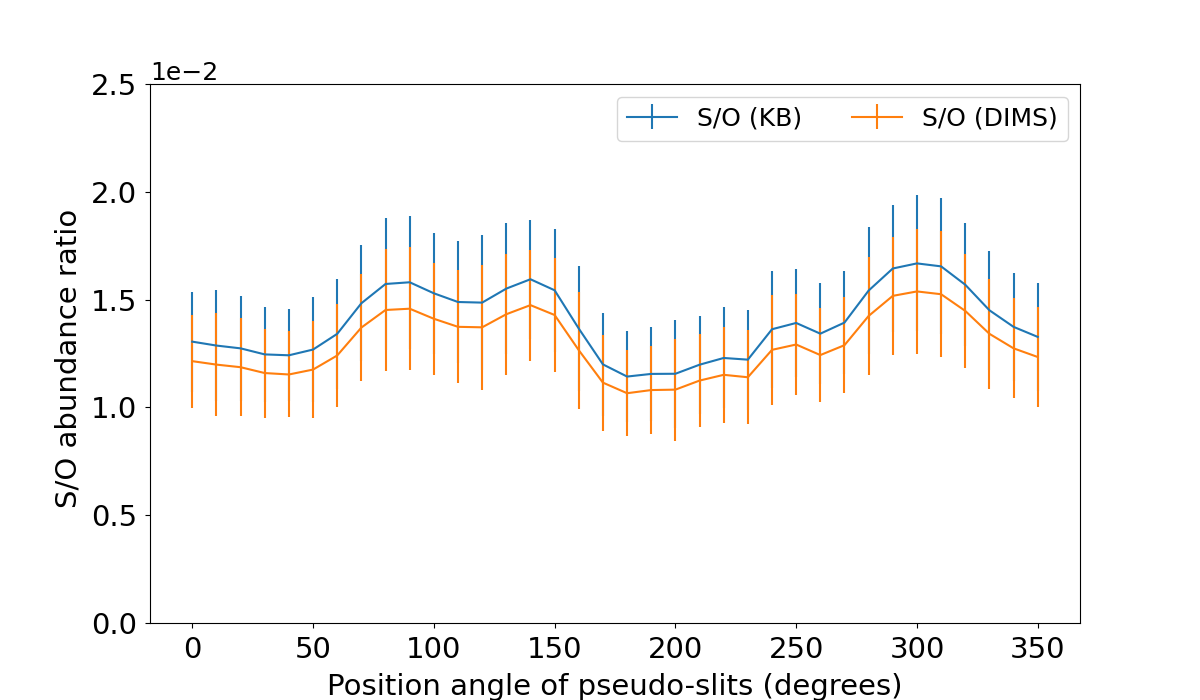}
\caption{Ionic and total S (upper panel) abundances and S/O ratio (lower panel) of NGC~6778 a function of the pseudo-slit position angles. The total abundances is computed using the ICFs schemes from \protect\cite{Kingsburgh1994K} and \protect\cite{Delgado2014}. The ionic abundances derived from \sulfurt~$\lambda$6716 and $\lambda$6731, and \sulfur~$\lambda$6312 and $\lambda$9069 emission lines are very similar which makes one overlays the other.} 
\label{figOSabundPA6778}
\end{figure} 

Fig.~\ref{figregions6778} illustrates the 10 pseudo-slits/regions selected for the spectroscopic exploration. The magenta slit (number 1) corresponds to the slit position from \cite{Dufour2015}\footnote{Our pseudo-slit is 0.6~arcsec (3~pixel) wide while the real one is 0.5~arcsec wide. Moreover, \cite{Dufour2015} excluded a small region from -0.36 to 0.72 arcsec to avoid contamination from the central star and the spectrum is the result of the sum of two extracted windows (-3.24:-0.36 and 0.72:3.6). {\sc satellite} calculates the spectra from the entire pseudo-slit (length=6.84~arcsec).}. The green pseudo-slit~2 has the same width as the pseudo-slit~1, but it is 60~arcsec longer, in order to cover the nebula from one side to the other and compare the results with those from the pseudo-slit~1 centred only on the inner zone. The cyan pseudo-slit~3 corresponds to the slit of \cite{Jones2016}\footnote{The {\sc satellite} pseudo-slit  and the one from \cite{Jones2016} do not correspond to the exact same regions. In particular, our pseudo-slit is slightly wider than the real one. It has a width of 0.8~arcsec (4~pixel) while the real one is 0.7~arcsec wide. Moreover, there was an offset between the red and blue exposures in \citet{Jones2016}, which we do not take into account.} (PA=$-$55~degrees). Pseudo-slits~4 to 9 (green colour) cover a number of LISs distributed throughout the nebula. Finally, pseudo-slit~10 (blue colour) covers the whole central region of NGC~6778.

In Table~\ref{fluxlinesPN_NGC6778}, we present the interstellar extinction corrected line intensities for the pseudo-slits as well as the integrated spectrum for an area of 1602~square arcsec. The spectra from the pseudo-slits~1 and 3 are compared with the observations in \cite{Dufour2015} and \cite{Jones2016}, respectively, and a very good agreement is found for the majority of the lines and physical parameters (c(\hb), T$_{\rm e}$ and N$_{\rm e}$). The integrated F(\hb) flux of the whole nebula is being 5.72$\times$10$^{-12}$~erg~s$^{-1}$~cm$^{-2}$, in reasonable agreement with the value of 6.92$\times$10$^{-12}$~erg~s$^{-1}$~cm$^{-2}$ found by \cite{Acker1992}.

The extinction coefficient is found to be 0.65 with a standard deviation of 0.02 between the values reported by \cite{Dufour2015} and \cite{Jones2016}. T$_{\rm e}$\sulfur~and N$_{\rm e}$\sulfurt~measured for the pseudo-slits vary from 8300 to 8600~K and from 550 to 850~cm$^{-3}$, respectively, in agreement with the literature. 

Chemical abundances are provided only by \cite{Jones2016} and can be compared with those obtained for the pseudo-slit~3. Despite the agreement in the line intensities and T$_{\rm e}$/N$_{\rm e}$, a significant difference in the chemical abundances is found. In particular, we find slightly higher He, twice higher O and S, half N and comparable or half Ar abundances depending on the ICF. This discrepancy in abundances is associated to a combination of the different extinction, physical conditions, atomic data sets and ICFs adopted. A more detailed analysis of such differences has been presented in \citet{Garciarojas2022}. All chemical abundances derived from the pseudo-slits and the central region (Table~\ref{fluxlinesPN_NGC6778}) show small variations, within their uncertainties.

\begin{table*}
\caption{Optical emission line intensities from the specific slit task of NGC~6778} \label{fluxlinesPN_NGC6778}
\begin{tabular}{lcccccccccccc}
\hline

Ion              &   I$_{S_D)}$& I$_{D}$ & I$_{S_D}~\rm{ext^{\dag}}$  & I$_{S_J}$ & I$_{J}$ & I$_{S}$ & I$_{S}$ & I$_{S}$ & I$_{S}$& I$_{S}$ & I$_{S}^{\dag\dag}$\\
                     & p-slit~1 & slit~1 & p-slit 2 & p-slit 3 & slit~3 & p-slit~4 & p-slit~6 & p-slit~8 & p-slit~9 & p-slit~10 & 1602 arc$^2$\\
\hline
He~{\sc ii}~4686~\AA\ &  17.4  &  19.8  &   8.26 &   4.07  &   6.50 &  13.4   &   1.66  &   0.624 &  11.4   &   5.69   &   5.10 \\
H~{\sc i}~4861~\AA\   & 100    & 100    & 100     & 100    & 100    & 100     & 100     & 100     & 100     & 100      & 100 \\ 
\oiiia                & 171    & 171    & 171     & 169    & 171.11 &  99.5   & 119     & 141     &  92.3   & 171      & 167 \\ 
\nia                  &   2.12 &   -    &   2.69  &   2.46 & -      &  15.8   &   1.60  &   3.50  &  13.5   &   2.99   &   3.15\\
\cliiia               &   0.548&   -    &   0.646 &   0.652& -      &   5.84  &   0.610 &   0.672 &   6.02  &   0.633  &   0.715\\
\cliiib               &   0.465&   -    &   0.556 &   0.553& -      &   6.08  &   0.489 &   0.557 &   5.65  &   0.544  &   0.615\\
\niia                 &   2.58 &   -    &   2.76  &   2.40 & -      &  11.2   &   1.34  &   2.39  &   9.98  &   2.91   &  2.87 \\
He~{\sc i}~5876~\AA\  &  23.0  &  26.1  &  23.2   &  23.4  &  22.12 &  27.0   &  23.7   &  23.3   &  25.2   &  23.3    &  23.4 \\
\oi                   &   3.34 &   -    &   4.41  &   3.81 &   3.69 &  32.1   &   1.66  &   5.15  &  24.9   &   4.97   &   5.05 \\
\siiia                &   1.09 &   -    &   1.19  &   1.11 &   1.05 &   4.34  &   0.695 &   0.905 &   2.64  &   1.21   &   1.20 \\
H~{\sc i}~6563~\AA\   & 296    & 286    & 296     & 296    & 289.88 & 297     & 297     & 297     & 296     & 297      & 296 \\
\niib                 & 195    & 153    & 257     & 229    & 226.65 & 879     & 142     & 302     & 981     & 282      & 279 \\
He~{\sc i}~6678~\AA\  &   6.57 &   8.54 &   6.65  &   6.72 &   6.61 &   9.41  &   6.79  &   6.64  &   7.90  &   6.69   &   6.71 \\
\siia                 &  14.3  &  12.9  &  19.3   &  17.1  &  16.92 &  59.7   &  12.1   &  24.78  &  64.3   &  20.8    &  21.1  \\ 
\siib                 &  16.3  &  14.7  &  21.0   &  18.1  &  18.00 &  44.9   &  11.1   &  23.7   &  48.9   &  23.1    &  22.6 \\
\ariii                &  14.6  &  11.0  &  16.8   &  16.4  & -      &  19.6   &  13.2   &  17.1   &  20.2   &  17.4    &  17.3 \\
\oiia                 &   3.14 &   -    &   2.58  &   2.30 & -      &   4.99  &   1.25  &   1.56  &   4.52  &   2.51   &   2.29\\
\oiib                 &   2.54 &   -    &   2.08  &   1.88 & -      &   8.39  &   1.13  &   1.26  &   4.99  &   2.04   &   1.20\\
\siiib                &  23.7  &  17.7  &  27.6   &  27.3  & -      &  18.1   &  21.6   &  22.8   &  20.5   &  28.4    &  27.8 \\
\hline
F(\hb)(10$^{-14}$)    & 10.7  & 3.09    & 27.6     & 26.6  &  -        & 0.05  & 0.79   &   5.77   &  0.04   & 465  &  572\\
c(\hb)                & 0.65  & 0.74    & 0.65     & 0.65  & 0.46      & 0.69  & 0.65   &   0.65   &  0.60   & 0.65 &  0.64\\
Te[N~{\sc ii}]        & 9524  & 10300$^a$& 8830    & 8769  & 8850$^b$  & 9441  & 8472   &   7996   &  8704   & 8689 &  8697\\
Te[S~{\sc iii}]       & 8578  &  -      &  8388    & 8219  & 8800$^b$  & 22454$^c$ & 7591   &   8139   & 13945$^c$   & 8329 &  8374\\
Ne[S~{\sc ii}]        &  781  &  919    &   661    &  586  &  590      &    90 &  318   &    380   &   100   & 685  &   610\\
Ne[Cl~{\sc iii}]      & 1179  &  -      &  1262    & 1156  &   -       &  3308 &  784   &   1014   &  2054   & 1249 &  1262\\
\hline        
\ch{He+}(5876)/\ch{H+}      & 0.161   &  -     &  0.161   & 0.163   & 0.154     & -     & 0.163  & 0.163    & -       & 0.162 & 0.163 \\
\ch{He+}(6678)/\ch{H+}      & 0.161   &  -     &  0.163   & 0.164   &  -        & -     & 0.164  & 0.162    & -       & 0.164 & 0.164\\
\ch{He^{++}}(4686)/\ch{H+}  & 0.014   &  -     &  0.007   & 0.003   & -         & -     & 0.013  & 0.001    & -       & 0.001 & 0.004\\
He/H                        & 0.175   &  -     &  0.169   & 0.167   & 0.159     & -     & 0.165  & 0.163    & -       & 0.167 & 0.168\\
\ch{N$^0$}/\ch{H+}~(-6)     & 7.13     & -      & 9.50      & 9.16      &  -     & -     & 7.42     & 12.2       & -   & 11.1      & 10.0\\
\ch{N+}/\ch{H+}~(-5)        & 5.99     & -      & 8.48      & 7.99      &  5.55  & -     & 6.37     & 1.09       & -   & 9.58      & 9.31\\
ICF(N)                      & 2.51/2.54& -      & 2.56/2.79 & 2.52/2.94 &  ?     & -     & 2.21/2.76& 2.62/3.71  & -   & 2.56/2.90 & 2.57/2.93\\
N/H~(-4)                    & 1.51/1.52& -      & 2.17/2.36 & 2.01/2.35 &  4.04  & -     & 1.41/1.76& 2.88/4.06  & -   & 2.45/2.77 & 2.39/2.73\\
\ch{O$^0$}/\ch{H+}~(-6)     & 12.0      & -      & 17.4      & 16.0      & -       & -    &  11.1     & 22.9      & -    & 20.1       & 20.2\\
\ch{O+}/\ch{H+}~(-4)        & 2.24      & -      & 2.26      & 2.33      & 0.56    & -    &  2.98     & 1.95      & -    & 2.29       & 2.18\\
\ch{O^{++}}/\ch{H+}~(-4)    & 3.08      & -      & 3.37      & 3.62      & 2.83    & -    &  3.58     & 3.15      & -    & 3.46       & 3.34\\
ICF(O)                      & 1.06/1.05 & -      & 1.03/1.02 & 1.01/1.01 & ?       & -    &  1.00/1.00& 1.00/1.00 & -    & 1.02/1.02  & 1.02/1.01\\
O/H~(-4)                    & 5.63/5.58 & -      & 5.78/5.76 & 6.14/6.12 & 3.39    & -    &  6.60/6.59& 5.11/5.11 & -    & 5.86/5.84 & 5.58/5.57\\
\ch{S+}/\ch{H+}~(-6)        & 1.33      & -      & 1.84      & 1.67      & 1.44    & -    & 1.34      & 2.29      & -    & 2.06       & 1.98\\
\ch{S^{++}}/\ch{H+}~(-6)    & 4.88      & -      & 5.99      & 6.21      & 4.26    & -    & 5.97      & 5.34      & -    & 6.27       & 6.06\\
ICF(S)                      & 1.09/1.02 & -      & 1.09/1.01 & 1.09/1.01 &  ?      & -    & 1.06/1.0  & 1.09/1.01 & -    & 1.09/1.0   & 1.09/1/00\\
S/H~(-5)                    & 0.68/0.0.63& -     & 0.85/0.78 & 0.86/0.78 & 0.34   & -    & 0.78/0.71 & 0.84/0.76 & -    & 0.91/0.83  & 0.88/0.80\\
\ch{Cl^{++}}/\ch{H+}~(-7)   & 1.04      & -      & 1.33    & 1.43        &   -     & -    & 1.75      &   1.51    & -    & 1.34       & 1.49\\
ICF(Cl)                     & -/1.37    & -      & -/1.34  & -/-         &   ?     & -    & -/-       &  -/-      & -    &  -/1.33    & -/1.33\\
Cl/H~(-7)                   & -/1.43    & -      & -/1.79  & -/-         &   -     & -    & -/-       &  -/-      & -    &  -/1.79    & -/1.97\\
\ch{Ar^{++}}/\ch{H+}~(-6)   & 1.78      & -      & 2.16     & 2.25      &  3.80    & -    & 2.28      &  2.39     & -    & 2.29      & 2.24\\
ICF(Ar)                     & 1.87/1.11 & -      & 1.87/1.10& 1.87/1.09 &   ?      & -    & 1.87/1.05 & 1.87/1.09 & -    & 1.87/1.09 & 1.87/1.09 \\
Ar/H~(-6)                   & 3.32/1.98 & -      & 4.05/2.36& 4.20/2.44 &  4.68    & -    & 4.26/2.39 & 4.47/2.61 & -    & 4.28/2.51 & 4.19/2.45 \\
\hline
\end{tabular}
\begin{flushleft}
I is the intensity of the lines. The indices S, D, and J 
point out the results obtained with {\sc satellite} or from previous studies: \citet{Dufour2015} and \citet{Jones2016}, $^{\dag}$ Extended pseudo-slit, $^{\dag\dag}$ Integrated from an area of 1602arc$^2$, $^a$: \cite{Dufour2015} use this value as default and do not provide ionic and total abundances, $^b$: T$_{\rm e}$ is derived from \sulfurt\ (row: T$_{\rm e}$[N~{\sc ii}]) and \oxygeniii~ (row:T$_{\rm e}$[S~{\sc iii}]) lines, $^c$: Very uncertain T$_{\rm e}$ values. Chemical abundances are not provided for this very faint pseudo-slits, \lq\lq ?\rq\rq=unknown value
\end{flushleft}
\end{table*}

The next step is the spectroscopic analysis of the pseudo-slit spectra as a function of the position angles (PA). c(\hb), T$_{\rm e}$[S~{\sc iii}] and N$_{\rm e}$[S~{\sc ii}] show no variability with PAs and their values are approximately 0.66, 8500~K and 680~cm$^{-3}$, with standard deviations of 0.01, 100~K and 80~cm$^{-3}$, respectively (Fig.~\ref{figTeNePA6778}). 

\begin{figure}
\centering
\includegraphics[width=7.8cm]{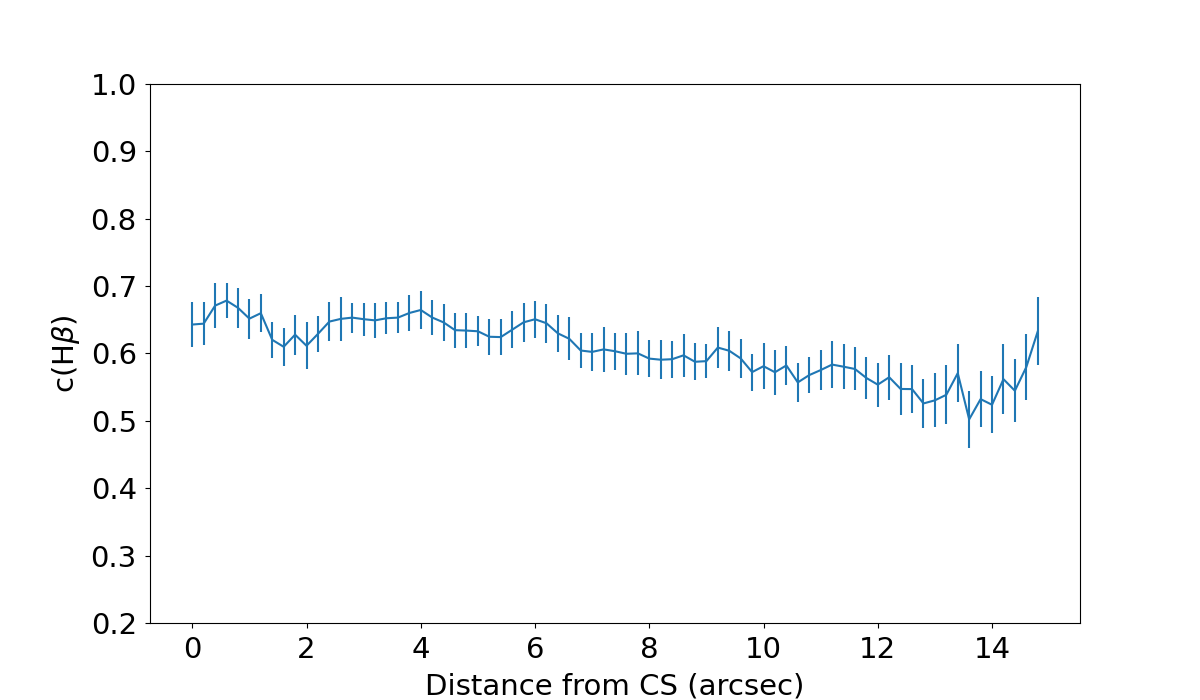}
\includegraphics[width=7.8cm]{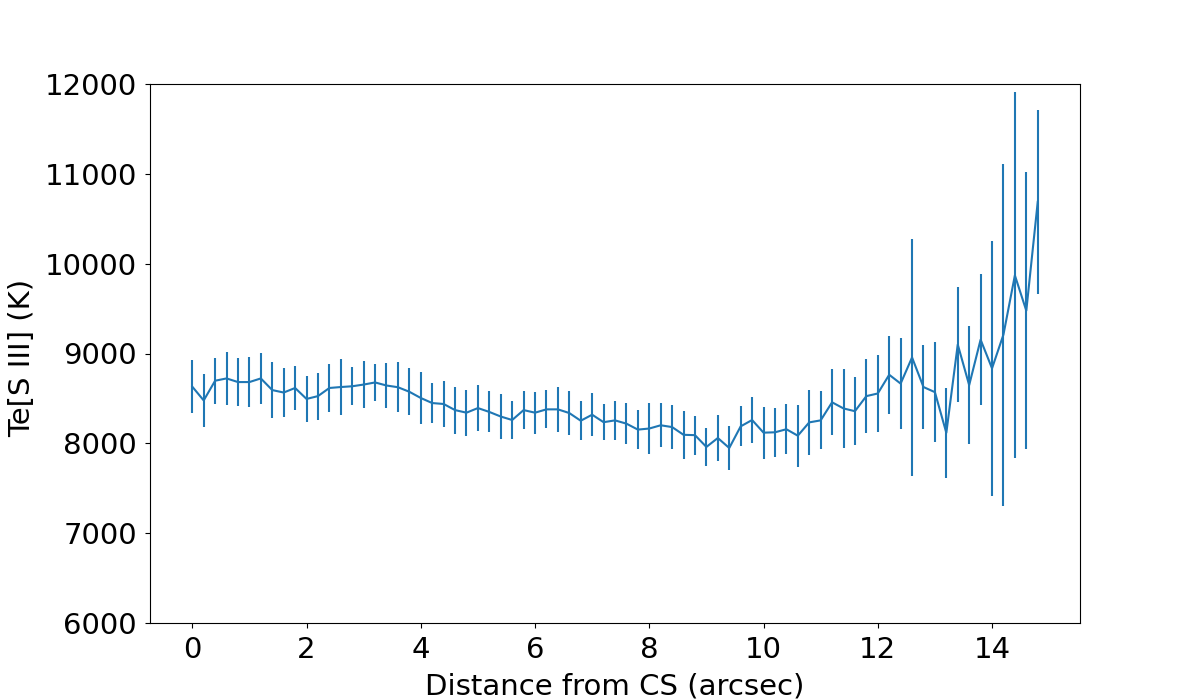}
\includegraphics[width=7.8cm]{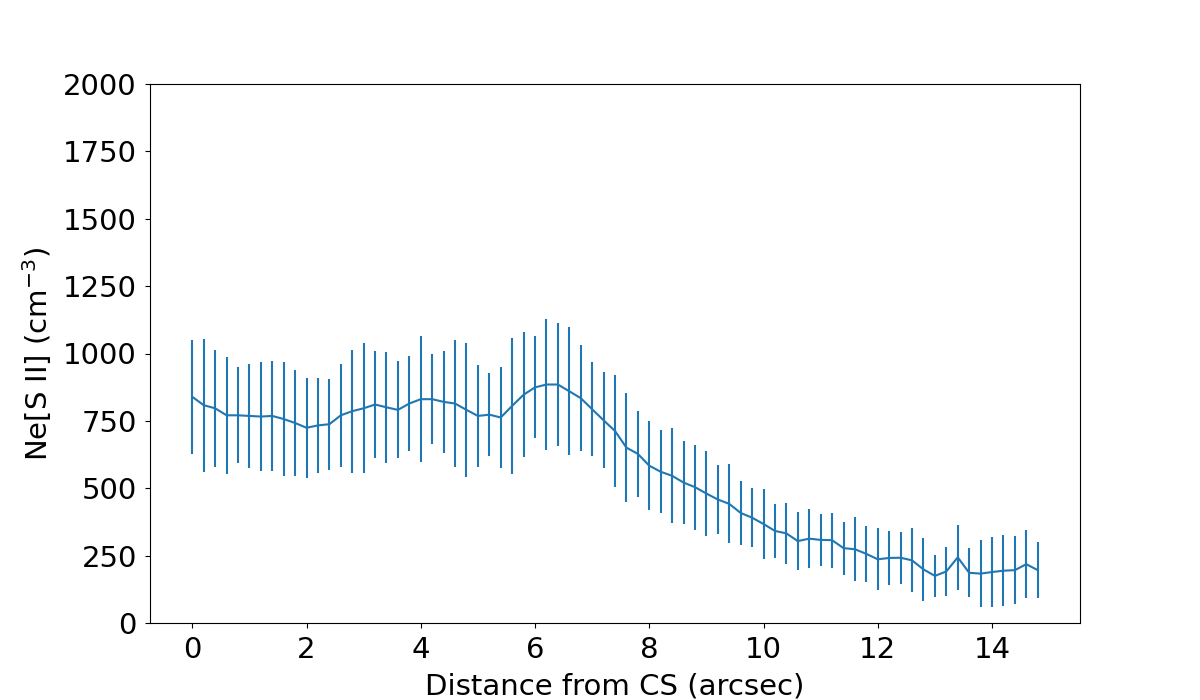}
\caption{c(H$\beta$), T$_{\rm e}$\sulfur~and N$_{\rm e}$\sulfurt~of NGC~6778 as functions of the distance from the central star in arcsec obtained from a pseudo-slit at PA=288~degrees.} 
\label{figTeNedistance6778}
\end{figure}

Regarding the chemical abundances, He and O appear unchanged with PAs, while N and S show an enrichment at 80$<$PA$<$150 and 260$<$PA$<$340~degrees (see Figs~\ref{figONabundPA6778} and \ref{figOSabundPA6778}). The N/O and S/O abundances ratios also appear to vary with PAs but they can be considered constant at $\sim$0.35 and $\sim$0.014 within the errors, respectively. This specific range of PAs, with enhanced N and S abundances, does not correspond to any particular knot or jet in NGC~6778, rather the equatorial waist of the nebula which is known to displays a filamentary wisp structure bright in \nitrogen~$\lambda$6584 line \citep{Guerrero2012}.

In order to further investigate the issue just discussed, the radial analysis of NGC~6778 as a function of the distance from the central star is performed for a pseudo-slit at PA=288~degrees across the equatorial waist (Fig.~\ref{figTeNedistance6778}). No variation in the c(\hb) is found with the distance from the CS up to 15~arcsec. T$_{\rm e}$ is nearly constant ($\sim$8500~K) up to 10~arcsec and then starts increasing outwards. However, the high uncertainties do not allow to argue whether this increment of T$_{\rm e}$ for distances $>$10~arcsec is real, due to an extra heating mechanism. N$_{\rm e}$ is also constant ($\sim$800~cm$^{-3}$) for a distance up to $\sim$7~arcsec and then smoothly decreases for larger distances down to 150/200~cm$^{-3}$.

\begin{figure}
\centering
\includegraphics[width=8.5cm]{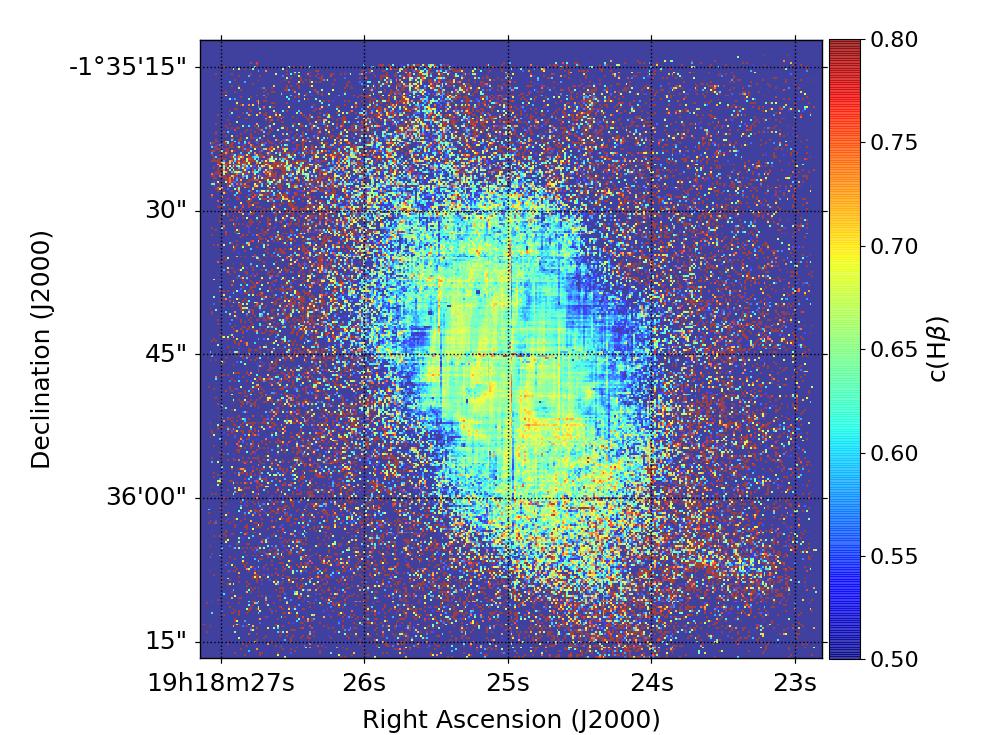}
\includegraphics[width=8.5cm]{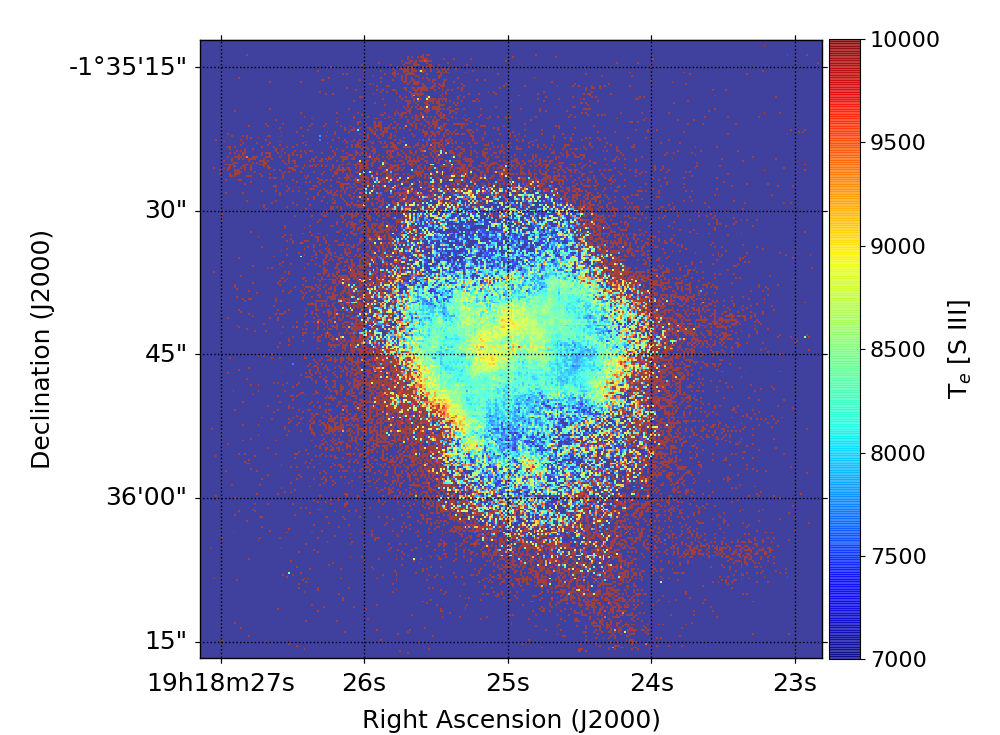}
\includegraphics[width=8.5cm]{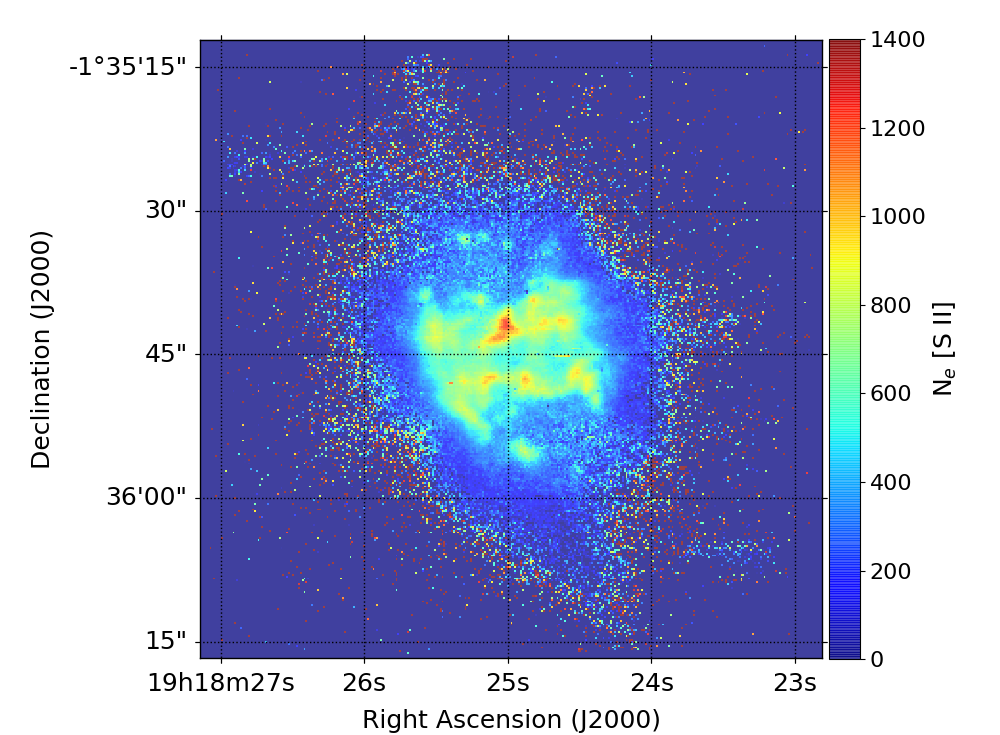}
\caption{\rm{c(H$\beta$) (upper), $T_{e}$[S~{\sc iii}] (middle), $N_{e}$[S~{\sc ii}] (lower)} maps of NGC~6778.}
\label{TeNeimage_ngc6778}
\end{figure}

The final part of the analysis of NGC~6778 is carried out through the 2-dimensional line maps which provide a better illustration of the spatial distribution for the nebular parameters. Fig.~\ref{TeNeimage_ngc6778} displays the maps of c(\hb), T$_{\rm e}$[S~{\sc iii}] and N$_{\rm e}$[S~{\sc ii}]. c(\hb) varies between 0.5 and 0.8. The values higher than 0.75 at the outskirt of the nebula are uncertain due to the low S/N ratio while the values lower than 0.55 found in the external parts of the disrupted equatorial component are likely real. Moreover, the 1D analysis returns a nearly constant value of 0.65 regardless of the PA of the pseudo-slit and it is in excellent agreement with the integrated value of 0.64 obtained with {\sc satellite} \citep{Akras2022} and 0.66 from \citet{Garciarojas2022}. A small spatial variation in T$_{\rm e}$[S~{\sc iii}] is observed throughout the nebula. The central region is characterized by a higher temperature of $\sim$9000~K while the rest of the nebula exhibits T$_{\rm e}$[S~{\sc iii}]$\sim$8500~K. On the other hand, N$_{\rm e}$[S~{\sc ii}] map shows evident density clumps (900$<$N$_{\rm e}<$1200~cm$^{-3}$) distributed in a more tenuous central region with N$_{\rm e}\sim$600~cm$^{-3}$.

\begin{figure}
\centering
\includegraphics[width=8.5cm]{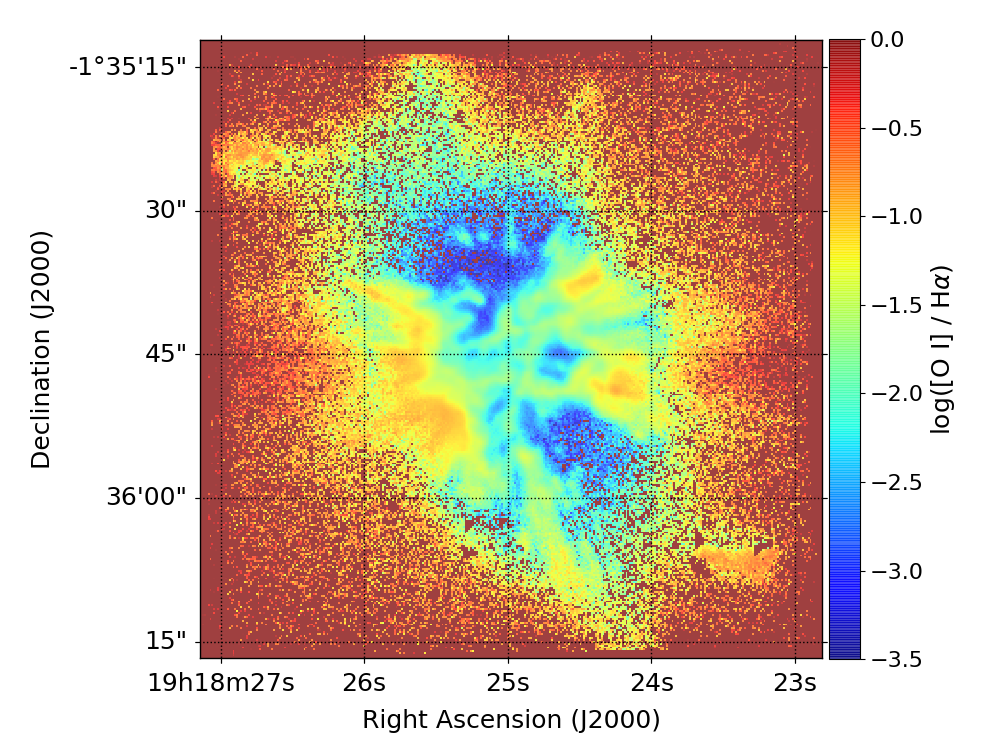}
\caption{A representative example of a line ratio map of NGC~6778: log(\oxygeni/\ha).}
\label{logimage_ngc6778}
\end{figure}

As a representative example of a line ratio map provided by {\sc satellite},  in Fig.~\ref{logimage_ngc6778} we present the log(\oxygeni~6300/\ha) map. The regions/knots 4,5 and 9 (see Fig.\ref{figregions6778}) are highlighted by this line ratio with value $>$-1. The outer parts of the equatorial waist ($>$8-9~arcsec) also display higher \oxygeni~6300/\ha~ratio relative to the central part of the nebula ($\sim$-2).

\begin{figure*}
\centering
\includegraphics[width=13.1cm]{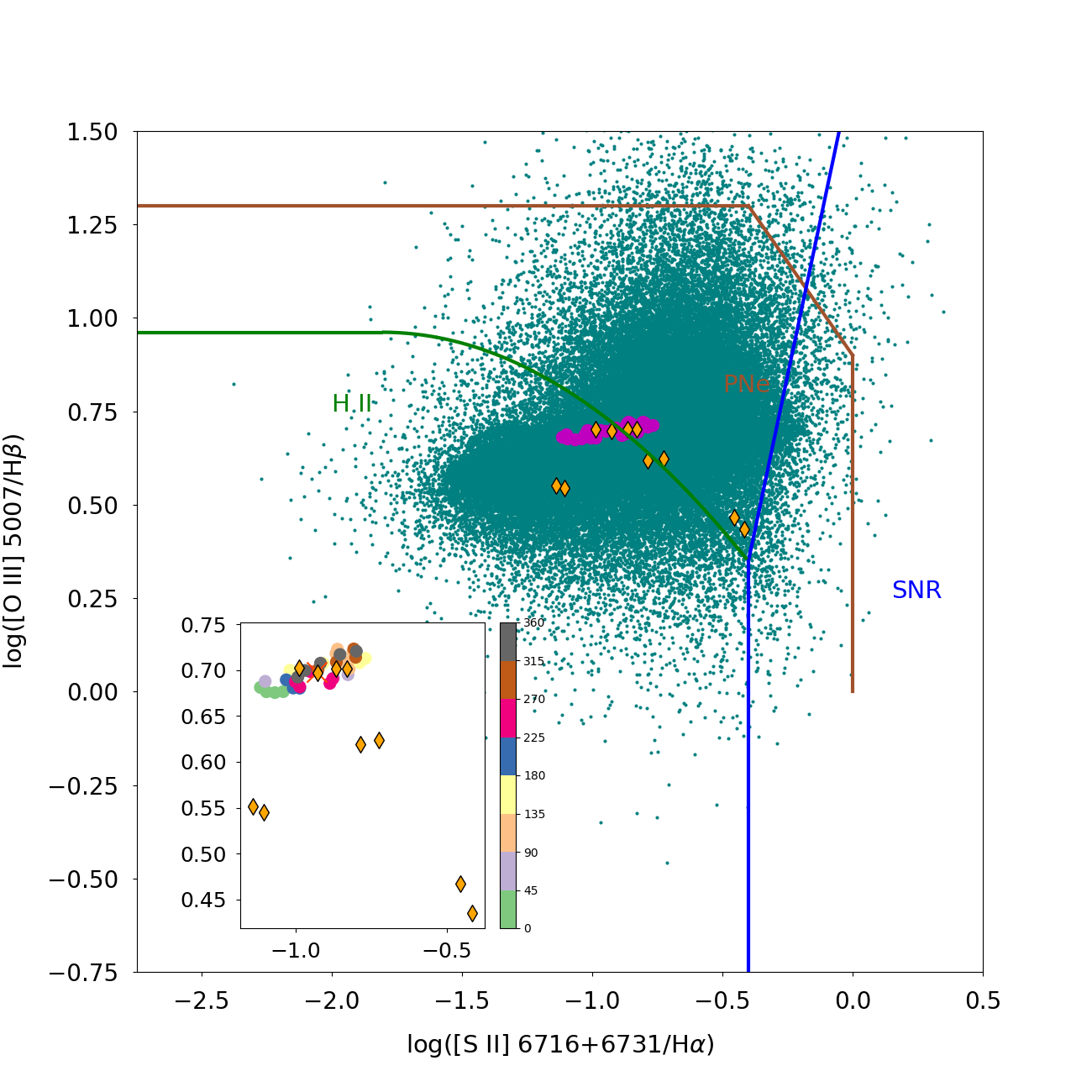}
\caption{Log(\sulfurt~6716+6731/\ha) versus Log(\oiiib/\hb) diagnostic diagram of NGC~6778 as a representative example. The symbols are the same as in Fig.~\ref{figHaSIIHaNII_DD}. The PNe, HII and SNRs zones are defined in \protect\citet{Frew2010,Sabin2013}.}
\label{figHaSIIHaNII_DD_ngc6778}
\end{figure*}

The emission line diagnostic diagram, extracted by {\sc satellite}, of the log(\oiiib/\hb) versus log(\sulfurt~6716+6731/\ha) is presented as a representative example of NGC 6778 (Fig.~\ref{figHaSIIHaNII_DD_ngc6778}). The nebula is UV dominated as indicated by the line ratios from the bulk of spaxel (blue dots), the {\it rotation} (purple circles) and {\it specific slits analysis} (yellow diamonds) modules are well distributed in the loci of PNe and H~{\sc ii} regions. It should be noted that two sub-structures of NGC~6778 display high log(\sulfurt~6716+6731/\ha) and log(\nitrogen~6548+6584/\ha) ratios, and they lie in a transition zone between the PNe and SNRs. These two sub-structures corresponds to the regions/pseudo-slit~4 and 9, two LISs at the end of the seemingly jet-like structures of NGC~6778 (see Fig.~\ref{figregions6778}). Scrutinising all the diagnostic diagram provided by {\sc satellite}, we conclude that these two sub-structures are distinct from the main nebula lying at the outer edges of the spaxels' distribution. An additional shock-heated mechanism cannot be ruled out. Finally, Table~\ref{statMUSEdatacube} lists the mean values and percentiles of c(\hb), $N_{e}$, $T_{e}$ and emission line ratios for the MUSE maps of NGC~6778.

\section{Discussion and conclusions} 

The transition from the traditional 1D long-slit spectroscopy to 2D imaging or integral field spectroscopy requires a proper interpretation of the results. To this end, a newly developed code written in {\sc python} language -- the Spectroscopic Analysis Tool for intEgraL fieLd unIt daTacubEs ({\sc satellite}) -- for the study of extended sources is presented. {\sc satellite} uses the 2D flux maps generated from the datacubes of any IFU to perform a full spectroscopic characterization of extended photoionized nebulae (PNe, H~{\sc ii} regions, galaxies, etc.) combining 1D and 2D analysis. 

In the current version of the {\sc satellite} code, four modules have been developed: {\it (I) rotation analysis, (II) radial analysis, (III) specific slits analysis and (IV) 2D analysis} modules. The MUSE Science Verification data of NGC~7009 was used for the verification of the performance of the code.

A comparison of the resulting 2D maps of several nebular parameters obtained with {\sc satellite} and those from \cite{Walsh2018} show a discrepancy less than 5\,per cent, except for the abundance of singly ionized oxygen (\ch{O+}/\ch{H+}) for which the difference is of $\sim$20~percent due to the recombination contribution to the \oxygenii~$\lambda\lambda$7320,7330 lines. In addition to the 2D maps, 1D spectroscopic analysis through 10 pseudo-slits resulted in a good agreement with previous long-slit spectroscopic studies. The evaluation of {\sc satellite's} outcomes was carried out using two different MUSE datasets of NGC~7009, verifying the efficiency and consistency of the code.

{\sc satellite} was also applied to the MUSE data of NGC~6778. A satisfactory agreement was found for the emission line intensities and physical parameters between {\sc satellite} and long-slit studies in the literature.

\begin{table*}
\caption{Statistical results of NGC~6778 MUSE datacube} \label{statMUSEdatacube}
\begin{tabular}{llrrrrrrr}
\hline
Parameter   & Num.~of~pix. & 5\% value &  25\% value &  50\% value & 75\% value &  95\% value & mean & SD  \\
\hline
c(\hb)                                 & 71074 & 0.11     & 0.43     &   0.61     & 0.67     & 1.21     & 0.60     & 0.31\\
Te(NII6548\_84)\_Ne(SII6716\_31)       & 30383 & 7500     & 8290     &   9230     & 10900    & 21400    & 11000    & 4270\\
Ne(SII6716\_31)\_Te(NII6548\_84)       & 35933 & 72       & 234      &   376      & 574      & 1760     & 582      & 571\\
Te(SIII6312\_9069)\_Ne(SII6716\_31)    & 25306 & 6830     & 7970     &   8490     & 9590     & 19500    & 10000    & 3820\\
Ne(SII6716\_31)\_Te(SIII6312\_9069)    & 25306 & 120      & 231      &   350      & 527      & 1070     & 488      & 550\\
Te(SIII6312\_9069)\_Ne(ClIII5517\_38)  & 13294 & 6950     & 7980     &   8340     & 8720     & 17300    & 9340     & 3220\\
Ne(ClIII5517\_38)\_Te(SIII6312\_9069)  & 13294 & 325      & 970      &   1770     & 2970     & 9280     & 2790     & 2700\\
\hline                                     
abundance~(He~{\sc i}~5876\AA)   &    25170 &  0.131    &  0.158    &    0.163    &  0.167    &  0.202    &  0.165    &  0.029 \\
abundance~(He~{\sc i}~6678\AA)   &    24921 &  0.122    &  0.159    &    0.164    &  0.170    &  0.245    &  0.173    &  0.054 \\
abundance~(He~{\sc ii}~4686\AA)  &    21188 &  1.92e-04 &  6.62e-04 &    1.98e-03 &  4.42e-03 &  1.75e-02 &  5.07e-03 &  1.05e-02 \\
abundance~(\oi)                  &    23444 &  1.29e-06 &  4.52e-06 &    1.19e-05 &  2.40e-05 &  7.83e-05 &  2.32e-05 &  3.53e-05 \\
abundance~(\oiia)                &    24296 &  6.44e-06 &  7.08e-05 &    1.90e-04 &  2.62e-04 &  8.46e-04 &  2.89e-04 &  5.60e-04 \\
abundance~(\oiib)                &    24402 &  9.38e-06 &  7.70e-05 &    1.93e-04 &  2.64e-04 &  9.32e-04 &  3.08e-04 &  6.25e-04 \\
abundance~(\oiiia)               &    25304 &  2.72e-05 &  1.47e-04 &    3.09e-04 &  3.74e-04 &  7.05e-04 &  3.18e-04 &  2.71e-04 \\
abundance~(\oiiib)               &    25304 &  2.69e-05 &  1.45e-04 &    3.06e-04 &  3.70e-04 &  6.96e-04 &  3.14e-04 &  2.68e-04 \\
abundance~(\nia)                 &    23377 &  1.29e-06 &  3.73e-06 &    7.75e-06 &  1.39e-05 &  4.08e-05 &  1.32e-05 &  1.76e-05 \\
abundance~(\niia)                &    23625 &  7.19e-06 &  3.68e-05 &    7.81e-05 &  1.24e-04 &  3.17e-04 &  1.20e-04 &  1.21e-04 \\
abundance~(\niic)                &    25052 &  8.34e-06 &  2.68e-05 &    5.75e-05 &  1.02e-04 &  2.51e-04 &  8.50e-05 &  7.80e-05 \\
abundance~(\niib)                &    25052 &  8.88e-06 &  2.77e-05 &    5.94e-05 &  1.05e-04 &  2.57e-04 &  8.75e-05 &  8.00e-05 \\
abundance~(\siia)                &    25306 &  2.38e-07 &  6.52e-07 &    1.28e-06 &  2.18e-06 &  5.42e-06 &  1.87e-06 &  1.75e-06 \\
abundance~(\siiia)               &    25306 &  1.26e-06 &  3.53e-06 &    5.31e-06 &  6.67e-06 &  9.72e-06 &  5.42e-06 &  2.77e-06 \\
abundance~(\cliiia)              &    22910 &  6.51e-08 &  1.12e-07 &    1.57e-07 &  2.19e-07 &  7.53e-07 &  2.50e-07 &  3.19e-07 \\
abundance~(\cliiib)              &    22415 &  7.16e-08 &  1.25e-07 &    1.77e-07 &  2.59e-07 &  9.30e-07 &  2.98e-07 &  3.93e-07 \\
abundance~(\ariii)               &    25268 &  3.83e-07 &  1.31e-06 &    2.03e-06 &  2.61e-06 &  3.97e-06 &  2.11e-06 &  1.23e-06 \\
\hline
log(He~{\sc i}~5876/\ha)                           & 66912 & -1.30  & -1.11  &  -1.08  & -0.96 & -0.50 & -0.99 & 0.34\\
log(He~{\sc ii}~4686/\hb)                          & 52019 & -2.40  & -1.48  &  -0.79  & -0.34 &  0.35 & -0.88 & 0.84\\
log(He~{\sc i}~5876/He~{\sc ii}~4686)              & 48946 & -0.67  & -0.16  &   0.29  &  0.76 &  1.79 &  0.38 & 0.92\\
log(\nitrogen~6584/\ha)                            & 71496 & -0.63  & -0.33  &  -0.75  &  0.05 &  0.35 & -0.12 & 0.30\\
log((\nitrogen~6548,6584)/\nitrogen~5755)          & 56399 &  0.62  &  1.10  &   1.65  &  2.02 &  2.26 &  1.56 & 0.63\\
log((\nitrogen~6548,6584)/(\oxygeniii~4959,5007))  & 71047 & -0.79  & -0.56  &  -0.35  & -0.20 &  0.16 & -0.37 & 0.65\\
log(\nitrogena~5200/\hb)                           & 56165 & -2.06  & -1.44  &  -0.92  & -0.55 &  0.11 & -0.95 & 0.67\\
log((\sulfurt~6716,6731)/\ha)                      & 65251 & -1.40  & -1.09  &  -0.80  & -0.64 & -0.35 & -0.84 & 0.33\\
log(\sulfurt~6716/\sulfurt~6731)                   & 65250 & -0.37  & -0.05  &   0.04  &  0.10 &  0.38 &  0.03 & 0.23\\
log((\sulfurt~6716,6731)/(\sulfur~6312,9069))      & 45420 & -0.30  & -0.11  &   0.81  &  0.23 &  0.54 &  0.09 & 0.27\\
log(\oxygeni~6300/\ha)                             & 58687 & -2.59  & -1.80  &  -1.31  & -0.99 & -0.42 & -1.38 & 0.67\\
log((\oxygeni~6300,6363)/(\oxygeniii~4959,5007))   & 46717 & -2.69  & -1.99  &  -1.51  & -1.21 & -0.63 & -1.58 & 0.76\\
log((\oxygeni~6300,6363)/(\oxygenii~7320,7330))    & 37067 & -0.45  & -0.04  &   0.21  &  0.39 &  0.71 &  0.18 & 0.35\\
log(\oxygeniii~5007/\hb)                           & 71309 &  0.42  &  0.58  &   0.68  &  0.75 &  1.05 &  0.70 & 0.19\\
log((\oxygenii~7320,7330)/(\oxygeniii~4959,5007))  & 53580 & -2.42  & -2.23  &  -1.86  & -1.41 & -0.79 & -1.77 & 0.86\\
log((\chloro~5517,5538)/\hb)                       & 44263 & -2.04  & -1.81  &  -0.97  & -0.39 &  0.25 & -0.99 & 0.81\\
log(\chloro~5517/\chloro~5538)                     & 44355 & -0.49  & -0.14  &   0.05  &  0.16 &  0.53 &  0.03 & 0.31\\
\hline
\end{tabular}
\begin{flushleft}
50\% percentile corresponds to the median.
\end{flushleft}
\end{table*}

The most noteworthy results from the spectroscopic analysis of the PNe studied in this work are:

\begin{itemize}
  \item The good agreement between the pseudo-slit spectra and the observed ones from long-slit spectroscopy in the literature (1D analysis).
  \item The variation of the line flux (or line ratios) as a function of the distance from the central star of NGC~7009 and the clear offset between the low and moderate/high ionization emission line (1D radial analysis).
  \item The perceptible augmentation of c(\hb) and $T_{e}$ in the LISs of NGC~7009  but not of NGC~6778 (1D and 2D analysis).
  \item The non-negligible spatial variation in c(\hb), $N_{e}$, $T_{e}$ and emission line ratios throughout both nebulae (2D analysis).
  \item The clear dispersion of sub-structures/features in the emission line diagnostic diagrams associated either with their ionization structure or the excitation mechanism (1D and 2D analysis).
  \item A spaxel-by-spaxel analysis of IFUs datacubes cannot be applied to emission line ratio diagnostic diagrams for distinguishing photo-ionized and shock-heated regions in resolved PNe (1D and 2D analysis).
  \item NGC~6778 shows considerable ionic abundance variations in N and S with the PAs of the pseudo-slits probably due to the ICFs, being enhanced in the direction of the equatorial waist (1D and 2D analysis). However, this variation is lessened in the abundances ratios.
  \item A discrepancy of $\sim$0.3~dex in the chemical abundances of NGC~6778 is found between the pseudo-slit 3 and the spectrum from \citet{Jones2016} is probably associated to the strong recombination affecting of the auroral \oxygeniii~$\lambda$4363 line. Although, our analysis is based on the  $T_{e}$(\sulfur), the diagnostic that may not be much affected by strong recombination emission.
  The atomic databases may also be responsible for this discrepancy. In particular, for NGC~7009 and NGC~6778, {\sc satellite} used the atomic databases (PYNEB\_18\_01) while \citet{Jones2016} used the {\sc chianti} database\footnote{https://www.chiantidatabase.org/}. \cite{Morisset2020} discuss the effect of different atomic data in the calculations of nebular parameters with an agreement of the order of 10~percent in the temperature sensitive line ratios and of 0.3~dex in chemical abundances (1D analysis).
\end{itemize}

Besides extended PNe, the {\sc satellite} code is also applicable to galaxies and H~{\sc ii} regions observed with any IFU as well as to the emission line maps obtained from 3D photo-ionization models such as the {\sc mocassin} code \citep{Ercolano2003,Ercolano2005} or the pseudo-3D models using the {\sc cloudy} code \citep{Ferland2013,Ferland2017} and the {\sc python} library {\sc pyCloudy} \citep{Morisset2013}. A full spectroscopic analysis of a source with all four models is doable in a day depending on the IFU and the size of the maps (number of spaxels.)

\subsection{Future implementations}

Further modules are planned to be implemented in the future version of {\sc satellite}. 

\begin{itemize}
    \item Faint collisionally excited lines as well as recombination lines, which may affect collisional lines sensitive to $T_{e}$ and $N_{e}$ measurements (such as \oxygeniii~4363 and \nitrogen~5755\AA).
    \item Chemical abundances will also be calculated from the recombination lines as well as the corresponding abundance discrepancy factor (ADF) for all the modules.
    \item A new module for non-rectangular regions (e.g. pseudo-slit) or complex regions/list of spaxels based on the flux/intensity of particular emission lines or line ratios will be included.
    \item Strong line methods for the determination of chemical abundances.
    \item The Voronoi tesselation method so that the {\sc satellite} code will be able to provide maps of emission line and physical parameters through an automatic segmentation based on the signal-to-noise ratio \citep{Vavilova2020}. This method will make possible the investigation of the faint halos around PNe \citep{Walsh2018}.
    \item The publicly available grid of photo-ionization ({\sc cloudy}) and shock models ({\sc mappings}) from the Mexican Million Models database \citep[3MdB,][]{Morisset2015,Alarie2019} will be employed in order to construct more general emission line diagnostic diagrams \citep[e.g. ][]{Akras2020a} and explore the contribution of each mechanism (photo-ionization and shocks).
    \item Principal component analysis tomography (PCA) technique will also be implemented to explore the variance in the emission of the observed source and reduce the dimentionality of the maps without losing much information.
    \item Near-infrared wavelengths are also planned to be implemented for the spectroscopic analysis of data from IFUs such as the Gemini-North's Near-Infrared Integral Field Spectrometer (NIFS@Gemini), the Spectrograph for INtegral Field Observations in the Near Infrared (SINFONI@VLT), the inFRared Imager and Dissector for Adaptive optics (FRIDA@GTC),  or the upcoming NIRSpec unit on the James Webb Space Telescope (JWST), the GMT integral field spectrograph (GMTIFS) among others.
\end{itemize}

\section*{Acknowledgements}
This paper is based on observations made with ESO Telescopes at the Paranal Observatory under Science Verification (SV) observing proposal 60.A-9347(A) and program ID 097.D-0241. SA acknowledges support under the grant 5077 financed by IAASARS/NOA. JG-R acknowledges support from the Severo Ochoa excellence program CEX2019-000920-S, from the State Research Agency (AEI) of the Spanish Ministry of Science, Innovation and Universities (MCIU) and the European Regional Development Fund (FEDER) under grant AYA2017-83383-P and from the Canarian Agency for Research, Innovation and Information Society (ACIISI), of the Canary Islands Government, and the European Regional Development Fund (ERDF), under grant with reference ProID2021010074. DGR acknowledges support from the CNPq grants 428330/2018-5 and 313016/2020-8. JG-R, DJ, and RC acknowledge support under grant P/308614 financed by funds transferred from the Spanish Ministry of Science, Innovation and Universities, charged to the General State Budgets and with funds transferred from the General Budgets of the Autonomous Community of the Canary Islands by the MCIU. DJ also acknowledges support from the Erasmus+ programme of the European Union under grant number 2020-1-CZ01-KA203-078200. This study was also financed in part by the Coordena\c{c}\~{a}o de Aperfei\c{c}oamento de Pessoal de N\'{i}vel Superior - Brasil (CAPES) - Finance Code 001 (IA). CM acknowledges support from grant UNAM / PAPIIT - IN101220. The following software packages in Python were used: Matplotlib \citep{Hunter2007}, NumPy \citep{Walt2011}, SciPy \citep{SciPy2020}, seaborn \citep{Waskom2021}, and AstroPy \citep{Astropy2013,Astropy2018}.

\section*{DATA AVAILABILITY}
The MUSE SV data underlying this article are available at the ESO online Archive (http://archive.eso.org/cms.html). The {\sc satellite code,} along with its documentation and examples, is available from the GitHub repository [https://github.com/StavrosAkras/SATELLITE.git].




\bibliographystyle{mnras}
\bibliography{references} 



\appendix

\bsp	
\label{lastpage}
\end{document}